\documentclass{aa}  
\usepackage{ulem}
\usepackage{graphicx}
\usepackage{txfonts}
\usepackage{bm}
\usepackage{color}
\makeatletter
 \def\@textbottom{\vskip \z@ \@plus 4pt}
 \let\@texttop\relax
\makeatother

\begin{document} 

   \title{Building protoplanetary disks from the molecular cloud: redefining the disk timeline.}

%   \subtitle{.}

   \author{K. Bailli\'e
          \inst{1,2}
          \and
          J. Marques\inst{3}%\fnmsep\thanks{Just to show the usage of the elements in the author field}
          \and
          L. Piau%\inst{2}%\fnmsep\thanks{Just to show the usage of the elements in the author field}
          }

   \institute{IMCCE, Observatoire de Paris, PSL Research University, CNRS,  Sorbonne Universit\'es, UPMC Univ Paris 06, Univ. Lille, 77 Av. Denfert-Rochereau, 75014 Paris, France
         \and
Centre National d'Études Spatiales, 2 place Maurice Quentin, 75039 Paris Cedex 01, France
%               Laboratoire AIM-LADP, Universit\'e Paris Diderot/CEA/CNRS, 91191 Gif sur Yvette, France.\\
        \and
Univ. Paris-Sud, Institut d’Astrophysique Spatiale, UMR 8617, CNRS, Batiment 121, 91405, Orsay Cedex, France
%               IMCCE, Observatoire de Paris, PSL Research University, CNRS,  Sorbonne Universit\'es, UPMC Univ Paris 06, Univ. Lille, 77 Av. Denfert-Rochereau, 75014 Paris, France.\\
%              \email{kevin.baillie@obspm.fr}
             }

   \date{Received August, 2018}

% \abstract{}{}{}{}{} 
% 5 {} token are mandatory
 
  \abstract
  % context heading (optional)
  % {} leave it empty if necessary  
{Planetary formation models are necessary to understand the characteristics of the planets that are the most likely to survive. Their dynamics, their composition, and even the probability of their survival depend on the environment in which they form. We therefore investigate the most favorable locations for planetary embryos to accumulate in the protoplanetary disk: the planet traps.}
  % aims heading (mandatory)
   {We study the formation of the protoplanetary disk by the collapse of a primordial molecular cloud, and how its evolution leads to the selection of specific types of planets.}
  % methods heading (mandatory)
   {We use a hydrodynamical code that accounts for the dynamics,
thermodynamics, geometry, and composition of the disk to numerically model its evolution as it is fed by the infalling cloud material. As the mass accretion rate of the disk onto the star determines its growth, we can calculate the stellar characteristics by interpolating its radius, luminosity, and temperature over the stellar mass from pre-calculated stellar evolution models. The density and midplane temperature of the disk then allow us to model the interactions between the disk and potential planets and determine their migration.}
  % results heading (mandatory)
   {At the end of the collapse phase, when the disk reaches its maximum mass, it pursues its viscous spreading, similarly to the evolution from a minimum mass solar nebula (MMSN). In addition, we establish a timeline equivalence between the MMSN and a "collapse-formed disk" that would be older by about 2 Myr.}
  % conclusions heading (optional), leave it empty if necessary 
   {We can save various types of planets from a fatal type-I inward migration: in particular, planetary embryos can avoid falling on the star by becoming trapped at the heat transition barriers and at most sublimation lines (except the silicates one). One of the novelties concerns the possible trapping of putative giant planets around a few astronomical units from the star around the end of the infall. Moreover, trapped planets may still follow the traps outward during the collapse phase and inward after it. Finally, this protoplanetary disk formation model shows the early possibilities of trapping planetary embryos at disk stages that are anterior by a few million years to the initial state of the MMSN approximation.}


\defcitealias{baillie14}{BC14}
\defcitealias{baillie15}{BCP15}
\defcitealias{baillie16}{BCP16}

   \maketitle
%
%________________________________________________________________

%%%%%%%%%%%%%%%%%%%%%%%%%%%%%%%%%%%
%1111111111111111111111111111111111
\section{Introduction} \label{intro}

The huge diversity of observed exoplanets is a challenge for planetary formation scenarios, which must not only explain the trends in the distribution of the exoplanet orbital periods and masses but also retrieve the observational constraints of the solar system. The observations of protoplanetary disks also provide additional constraints necessary to model the environment  and the conditions in which planets may form. This environment is the key element that determines which planets will fall onto their host star by spiraling inward by type-I migration in a few hundred thousand years \citep{goldreich79, arty93,ward97} and which ones will avoid that by becoming trapped at the density gradient discontinuities \citep{masset06b,paarp09a}, or at the opacity transitions \citep{menou04}. These latter transitions potentially result from the sublimation lines of the disk dust species (\citet{baillie15} and \citet{baillie16},  hereafter referred to as \citetalias{baillie15} and \citetalias{baillie16}). These planet traps also favor the accumulation of planetary embryos at certain radial distances from the star, where they may grow thanks to collisions \citep{morbidelli08}. Timescales are critical here since forming a planet by gas accretion on a solid core requires a few million years \citep{pollack96}, while the gas of the disk dissipates on a similar timescale \citep{font04, alexander07, alexander09, owen10}, as confirmed by disk observations by \citet{beckwith96} and \citet{hartmann98}.

Previous studies of planet migration mainly relied on simplified protoplanetary disk structures, mostly following power-law profiles as suggested by the classical minimum mass solar nebula (MMSN) \citep{weiden77, hayashi81}. \citet{hasegawa111} and \citet{paar11} used a similar framework to model the migration of planets in disks whose surface mass density and midplane temperature radial profiles followed power laws, while \citet{bitsch11} and \citet{bitsch13} used density prescriptions to reconstruct ($r,z$)-temperature maps. \citet{bitsch13} allowed the geometry to be consistently calculated with the thermal structure.

In the present article, not only do we model the evolution of the disk structure (in density and temperature) by setting its geometry free (\citet{baillie14}, hereafter \citetalias{baillie14}), but we also carry the initial state problem back from the usual MMSN to the primordial molecular cloud: we do so by modeling the formation of the disk itself by the gravitational collapse of the cloud, similarly to what \citet{hueso05} and \citet{yang12} suggested, but this time accounting for the growth of the star and the evolution of its physical properties. The viscous evolution of the disk is described in \citetalias{baillie14}, \citetalias{baillie15} and \citetalias{baillie16} and takes into account the shadowed regions that are not irradiated directly by the star, the variations of the dust composition of the disk with the temperature, and the evolution of the dust-to-gas ratio. In addition, we now take into consideration a proper model for the disk self-shadowing. We show that the disk tends to warm up during the collapse phase due to the stellar luminosity and that planet traps are carried away from the star, in particular at the sublimation lines (except for the silicates) and the heat transition barriers where the dominant heating process changes between viscous heating and stellar irradiation heating. We notice that a larger diversity of planets may be trapped than in the MMSN evolution models: in particular, these lines may hold Jupiter-like planets of a few hundred Earth masses located at a few astronomical units when the disk is so thick or viscous that this prevents gap opening. Finally, we highlight a migration mode dubbed "trapped migration" that allows planets to still migrate across the disk while remaining trapped, consistent with the results of \citet{lyra10}.

Section \ref{methods} describes how we numerically model the star+disk system: from the growth of the star by the collapse of the molecular cloud that also feeds the disk, to the viscous spreading of the gas and dust. We highlight the improvements of the code detailed in \citetalias{baillie14}, \citetalias{baillie15} and \citetalias{baillie16} that account in particular for the early stellar evolution, the self-shadowing of the disk, and the heating of the cloud. Based on the calculated evolution of the surface mass density and midplane temperature, Section \ref{res} aims to estimate the influence of the disk structure on the protoplanet migration, and to determine the position of the planet traps. Section \ref{disc} investigates how naturally evolving disk structures may be related to the survival of growing planets. Finally, Section \ref{cclpersp} summarizes our conclusions and details some perspectives that may provide a better understanding of the synthesis of the planet population.

%%%%%%%%%%%%%%%%%%%%%%%%%%%%%%%%%%%
%2222222222222222222222222222222222
\section{Methods} \label{methods}
\subsection{Disk evolution model}\label{dem}

Similarly to \citetalias{baillie14}, \citetalias{baillie15} and \citetalias{baillie16}, we model the protoplanetary disk as a viscous $\alpha$-disk \citep{shakura73} with a turbulent viscosity $\alpha_{\mathrm{visc}}= 10^{-2}$, as is usually taken for T Tauri-star disks without deadzones. Following \citet{lyndenbellpringle74} and \citet{pringle81}, we calculate the evolution of the disk surface mass density using the mass and angular momentum conservation. We assume that a cloud mass element $\mathrm{d}m_{\mathrm{infall}}$ joins the disk at radius $r$ over a time $\Delta t$, with an angular momentum $\mathrm{d}m_{\mathrm{infall}} \, r^2 \, \Omega(r)$. Due to the infall of material from the molecular cloud, the mass conservation reads

\begin{equation}
\frac{\partial \Sigma(r,t)}{\partial t} + \frac{1}{r}\, \frac{\partial}{\partial r}\left(r \, \Sigma(r,t) \, v_{r}(r,t) \right) =  S(r,t)
\label{massconservation}
,\end{equation}

with $\Sigma (r,t)$ being the surface mass density and $v_{r} (r,t)$ being the radial velocity of the gas in the disk. The left part is similar to the expression derived by \citet{pringle81} and the right part, $S(r,t)$, is a source term that accounts for the infall of the molecular cloud gas onto the disk at radius ${r}$. This source term relates to the infall mass element by $\mathrm{d}m_{\mathrm{infall}} = 2\, \pi \, r \, \Delta r \, S(r,t) \, \Delta t$.

The angular momentum conservation then reads

\begin{equation}
\begin{split}
r\, \frac{\partial}{\partial t}\, \left( r^2 \, \Omega \, \Sigma \right)  + &\frac{\partial}{\partial r}\, \left( r^2 \, \Omega \cdot r \, \Sigma \, v_{\mathrm{r}}\right) = \\
&\frac{\partial}{\partial r}\left(\nu \, \Sigma \, r^3 \, \frac{\partial \Omega}{\partial r} \right) + r \, S(r,t) \, r^2 \, \Omega,
\end{split}
\label{momentumconservation}
\end{equation}
with $\nu$ being the viscosity.

\begin{equation}
\begin{aligned}
r^3\, \Omega \frac{\partial \Sigma}{\partial t} + &r\, \Sigma \, \frac{\partial}{\partial t}\, \left( r^2 \, \Omega \right) + r^2 \, \Omega \, \frac{\partial}{\partial r}\, \left( r \, \Sigma \, v_{\mathrm{r}}\right) + r \, \Sigma \, v_{\mathrm{r}} \, \frac{\partial}{\partial r}\, \left( r^2 \, \Omega \right)\\
= &\frac{\partial}{\partial r}\left(\nu \, \Sigma \, r^3 \, \frac{\partial \Omega}{\partial r} \right) + r \, S(r,t) \, r^2 \, \Omega
\end{aligned}
.\end{equation}

While the term in $\frac{\partial}{\partial t}\, \left( r^2 \, \Omega \right)$ can usually be forgotten when the star only accretes mass from the disk at limited mass accretion rates, we cannot neglect it in this study as the star is gaining mass from both the molecular cloud (during the collapse phase) and the disk. In addition, Equation \ref{massconservation} allows to simplify the angular momentum equation:

\begin{equation}
r\, \Sigma \, \frac{\partial}{\partial t}\, \left( r^2 \, \Omega \right) + r \, \Sigma \, v_{\mathrm{r}} \, \frac{\partial}{\partial r}\, \left( r^2 \, \Omega \right) = \frac{\partial}{\partial r}\left(\nu \, \Sigma \, r^3 \, \frac{\partial \Omega}{\partial r} \right)
.\end{equation}

Assuming $\Omega \approx \Omega_{\mathrm{keplerian}}$, we can write

\begin{equation}
r \, \Sigma \, v_{\mathrm{r}} = -3\, \sqrt{r} \, \frac{\partial}{\partial r} \left( \nu \, \Sigma \, \sqrt{r}\right) - r^2 \, \Sigma \, \frac{\dot{M}_{*}}{M_{*}}
.\end{equation}

Finally, reporting this expression in equation \ref{massconservation}, we find Equation \ref{lb74}:

\begin{equation}
\frac{\partial \Sigma(r,t)}{\partial t} = \frac{3}{r} \, \frac{\partial}{\partial r}\left(\sqrt{r} \, \frac{\partial}{\partial r} \left( \nu(r,t) \, \Sigma(r,t) \, \sqrt{r}\right) \right) + S(r,t) + S_{2}(r,t)
\label{lb74}
,\end{equation}

where
\begin{equation}
S_{2}(r,t) = \frac{1}{r} \,  \frac{\partial}{\partial r}\left(r^2 \, \Sigma (r,t) \, \frac{\dot{M}_{*}}{M_{*}} \right)
.\end{equation}

Following \citet{hueso05} and \citet{yang12} and their discussions about the chosen approximations, we model the cloud envelope as isothermal and spherically symmetric and assume that the presolar cloud rotates rigidly with a constant angular velocity and that the cloud material collapses below the centrifugal radius (defined as the point at which the maximal angular momentum in the shell is equal to the angular momentum in the disk):
\begin{equation}
R_{\mathrm{c}}(t) = 10.5 \, \left(\frac{\omega_{\mathrm{cd}}}{10^{-14}\, \mathrm{s}^{-1}} \right)^{2} \, \left(\frac{T_{\mathrm{cd}}}{15 \, \mathrm{K}}\right) ^{-4}\, \left( \frac{M (t)}{1 \, M_{\odot}}\right)^{3} \mathrm{AU},
\label{rce}
\end{equation}
where $M (t)$ is the total mass of the star+disk system at instant $t$, and $\omega_{\mathrm{cd}}$ and $T_{\mathrm{cd}}$ are the rotational speed and temperature of the molecular cloud. The infall then happens in the inner parts of the disk, where $r < R_{\mathrm{c}}(t)$, and the source term is defined as

\begin{equation}
S(r < R_{\mathrm{c}}(t),t) = \frac{\dot{M}}{8\, \pi \, R_{\mathrm{c}}^{2}} \, \left( \frac{r}{R_{\mathrm{c}}} \right) ^{-3/2} \, \left[ 1 - \left(\frac{r}{R_{\mathrm{c}}}\right)^{1/2} \right] ^{-1/2}, 
\label{Srtm}
\end{equation}
with $\dot{M}$ being  the infall mass accretion rate from the cloud onto the star+disk system for which we use the expression of \citet{shu77}:
\begin{equation}
\label{Mdot}
\dot{M} = 0.975 \frac{c_{\mathrm{S}}^{3}}{G},
\end{equation}
where $c_{\mathrm{S}} = \sqrt{1.4 \, \frac{k_{\mathrm{B}}T_{\mathrm{cd}}}{\bar{m}}}$ is the cloud isothermal sound speed, $k_{\mathrm{B}}$ is the Boltzmann constant, and $\bar{m} = 2.3 \, m_{\mathrm{H}} \, = \, 3.8469 \, 10^{-27}$ kg, which is the mean molecular weight of the predominant $\mathrm{H}_{2}$ gas in the disk.

In line with the works of \citet{hueso05} and \citet{yang12}, we consider that the isothermal cloud provides material to the disk at a constant rate throughout the entire collapse phase given by Equation \ref{Mdot}. Finally, we set the collapse phase to end when the total mass of the star+disk system reaches $1 M_{\odot}$. The additional source term (Eq. \ref{Srtm}) is then canceled.

Among various attempts to model the collapse of the molecular cloud, \citet{mellon08} and \citet{hennebellef08} highlighted the limitations of the self-similar solution of \citet{shu77} based on a purely hydrodynamical model: their MHD simulations showed that the collapsing material can drag magnetic field inward, intensifying its strength and leading to an outward transport of angular momentum. This phenomenon of “magnetic braking” detailed in \citet{mellon08} can slow down the gas rotation, possibly preventing the formation of a centrifugally supported disk. Comparing numerical models of Class-0 protostars with observations from IRAM Plateau de Bure, \citet{maury10} showed that observations are better explained by magnetized models of protostar formation than by a purely hydrodynamical simulation in the absence of turbulence. However recent studies suggest a possible workaround to allow disk formation in a magnetic environment: while \citet{dapp10} suggested a reduction of the magnetic field strength by ohmic dissipation, \citet{joos12} investigated possible misalignments of the rotation axis and magnetic field direction, and \citet{seifried13} aimed at countering the magnetic braking by the disk turbulence. \citet{li14} provided an extensive review of the constraints and possible solutions for disk formation in the presence of a magnetic field, while \citet{tobin15} investigated the presence of such Class-0 disks in the Perseus region.
Though our model is less realistic from a magnetic point of view, it still considers the magnetic field as being related to the turbulent viscosity $\alpha$.

\citet{masunaga98} and \citet{masunaga99} showed that compressional heating might overtake radiative cooling and terminate the collapse phase. Though we believe that effect is of importance for the formation of the core, we chose to neglect that contribution for the disk, consistently with the models of \citet{hueso05} and \citet{yang12}. In addition, in our case, the termination of the collapse phase is determined by the initial mass of the cloud.

Similarly to \citetalias{baillie14}, \citetalias{baillie15} and \citetalias{baillie16}, we divide the disk into consecutive annuli that are logarithmically distributed in radius between $R_{*}$ and 1000 AU. We then model Equation \ref{lb74} numerically over a 1D grid of masses. The boundary conditions are set so that the disk cannot gain mass from the star at the inner edge. The mass accretion rate of the disk onto the star is therefore derived from the innermost mass flux.

For every radial position at every time, we model the thermodynamics and geometry of the disk by considering the heat equation described in Equations 15-18 from \citet{calvet91}, accounting for stellar irradiation heating, radiative cooling, viscous heating defined as $F_{v}(r) = \frac{9}{4} \Sigma(r) \nu(r) \Omega^{2}(r)$, and an external cloud envelope radiation heating $\sigma T_{\mathrm{cd}}^{4}$, with $\sigma$ being the Stefan-Boltzmann constant. Following the method thoroughly described in \citetalias{baillie14}, we solve this implicit equation
numerically on the disk midplane temperature $T(r)$. This allows us to derive the disk pressure scale height $h_{\mathrm{pr}}(r)$, its photosphere height $H_{\mathrm{ph}}(r)$, and its grazing angle $\alpha_{\mathrm{gr}}(r)$ (the angle of incidence of the stellar irradiation upon the disk photosphere) defined as

\begin{equation}
\alpha_{\mathrm{gr}}(r) = \arctan \left( \frac{dH_{\mathrm{ph}}}{dr}(r)\right) - \arctan\left(\frac{H_{\mathrm{ph}}(r)-0.4 R_{*}}{r}\right)
\label{alphagr}
.\end{equation}

Though the self-shadowing of the disk was not taken into account in \citetalias{baillie16}, we here consider that the irradiated parts of the disk cast shadows on the outer parts which have a lower incidence angle $\arctan(H_{\mathrm{ph}}(r)/r)$ than the shadowing inner parts. The value of the grazing angle determines whether a disk annulus is irradiated or not (in that case, we refer to "shadowed regions"). In addition, we consider the opacity model derived from \citet{helling00} and \citet{semenov03}, and detailed in \citetalias{baillie16} (see their Figure 1 and Table 1) to account for the composition of the dust species contained in the disk: water ice, volatile organics, refractory organics, troilite, olivine and pyroxene.

Since we aim at modeling the evolution of the disk over its lifetime and across wide radial scales, we are forced to neglect the heat diffusion. As we can see from the disks modeled by \citet{bitsch13}, \citet{bitsch14} and \citet{bitsch15a} that take into account the heat diffusion, we can expect the temperature structures to be smoothed radially. Therefore, the reader should keep in mind that our abrupt transitions (temperature plateau edges, heat transition barrier, optically thin frontier), should probably be smoother in real disks.

\subsection{Young star evolution model}
Using the "PHYVE" code (Protoplanetary disk HYdrodynamical Viscous Evolution) thoroughly detailed in \citetalias{baillie14}, \citetalias{baillie15} and \citetalias{baillie16}, we track the evolution of the mass distribution across the star+disk system: the viscous evolution governs the amount of material that is transferred from the disk to the star. In addition, the infall of mass onto the disk and onto the central star can be derived from Equations \ref{lb74} and \ref{Srtm}.

At every time step in the simulation, we interpolate the stellar radius $R_{*}$, luminosity $\mathcal{L}_{*}$ and temperature $T_{\mathrm{eff}}$ from tables of pre-calculated stellar evolutions modeled using the code CESAM detailed in \citet{morel97}, \citet{morel08} and \citet{piau11}, as well as \citet{marques13}.
These tables provide the radius, luminosity, and temperature of a constant-mass star as a function of its age and mass accretion rate. As these quantities are provided by the viscous evolution of the disk, we now have an empirical model for our star evolution, which is an interesting refinement compared to the fixed mass-accretion-rate star model used in \citet{hueso05}. For the purpose of the present study, we make the approximation that there is no accretion luminosity associated with the stellar growth. This is indeed a necessary simplification in order to use the pre-calculated stellar evolution tables. However, we are confident that this approximation only marginally affects the disk evolution, which is the main focus of the present study. Indeed, this accretion luminosity would only be of the same order of magnitude as the stellar luminosity during the collapse phase. At this time, a larger luminosity would result in a hotter disk and therefore in a faster viscous spreading of the disk. Since this only takes place in the first few hundred thousand years, this will only temporarily accelerate the growth of the star, and is very unlikely to affect the later evolution of the disk and star.

\subsection{Initial conditions}
While previous works from \citetalias{baillie14}, \citetalias{baillie15} and \citetalias{baillie16} considered an MMSN around a classical T Tauri-type star as the initial condition of their simulations, in the present paper we relax that assumption and model the early evolution of the star and the disk while the collapse of the primordial molecular cloud supplies them with gas and dust. However, as we do not seek to model the star ignition, we assume that the star has already formed and grown up to 0.2 $M_{\odot}$ at the start of our simulation.

In addition, we chose the initial molecular cloud to be consistent with the initial clouds of \citet{hueso05} and \citet{yang12}. Following \citet{vandishoeck1993}, who estimated the temperature of the cloud falls between 10 and 20 K, and consistently with \citet{yang12}, we chose the temperature of the cloud to be $T_{\mathrm{cd}} = 15$ K. We took its initial angular velocity $\omega_{\mathrm{cd}} = 10^{-14} \, \mathrm{s}^{-1}$, in accordance with the observed velocity gradients in the clouds by \citet{goodman93}, \citet{barranco98} and \citet{lodato08} that provide a range from $10^{-15}$ to $10^{-13} \, \mathrm{s}^{-1}$.

%%%%%%%%%%%%%%%%%%%%%%%%%%%%%%%%%%%
%3333333333333333333333333333333333
\section{Results} \label{res}

\subsection{Star and disk growth}

\begin{figure}[htbp!]
\begin{center}
\includegraphics[width=8cm, clip=true]{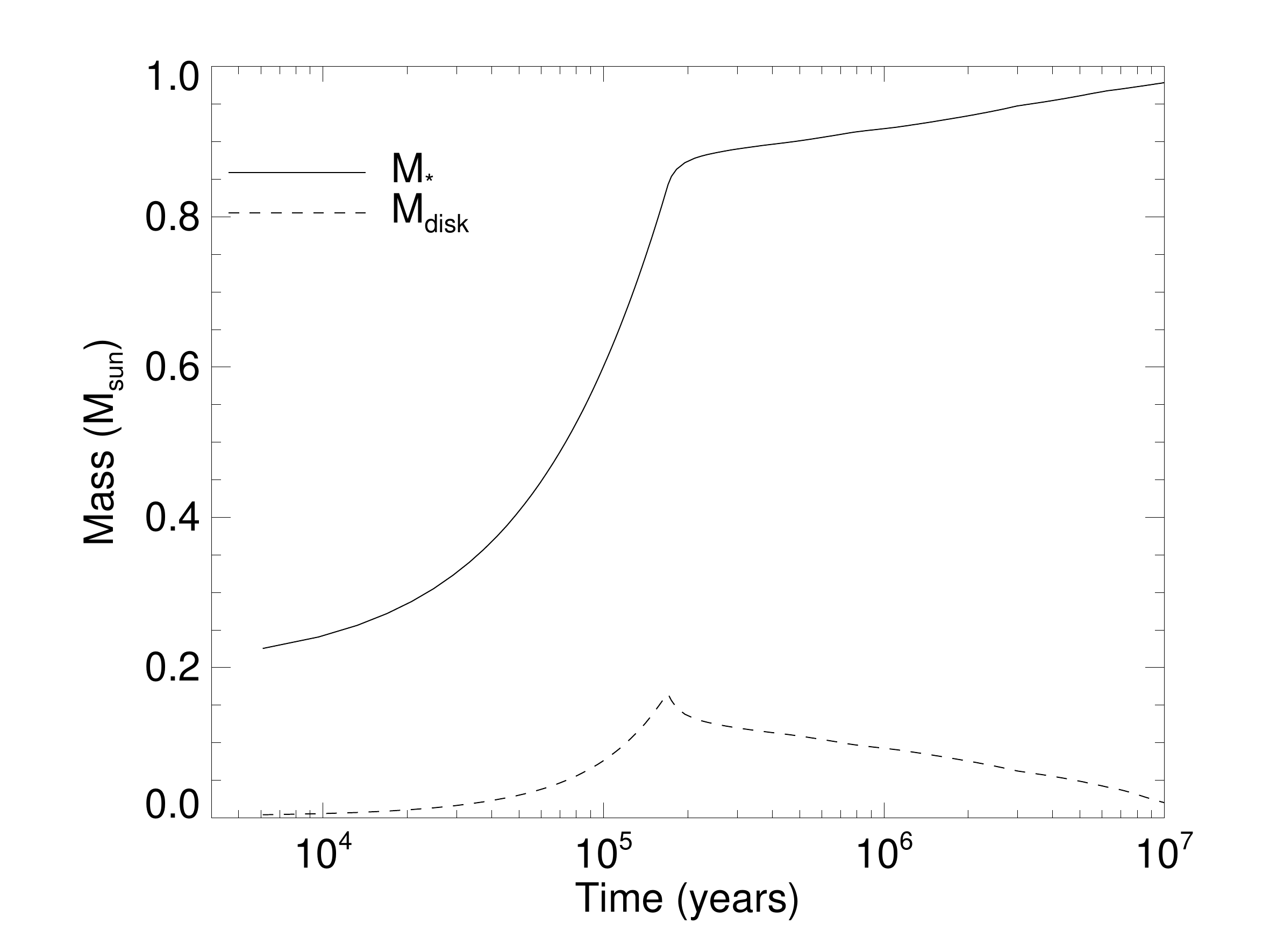}
\end{center}
\caption{Time evolution of the star and disk masses. The gravitational collapse that feeds the disk and the star ends after 170,000 years.}
\label{metoile}
\end{figure}

\begin{figure}[htbp!]
\begin{center}
\includegraphics[width=8cm, clip=true]{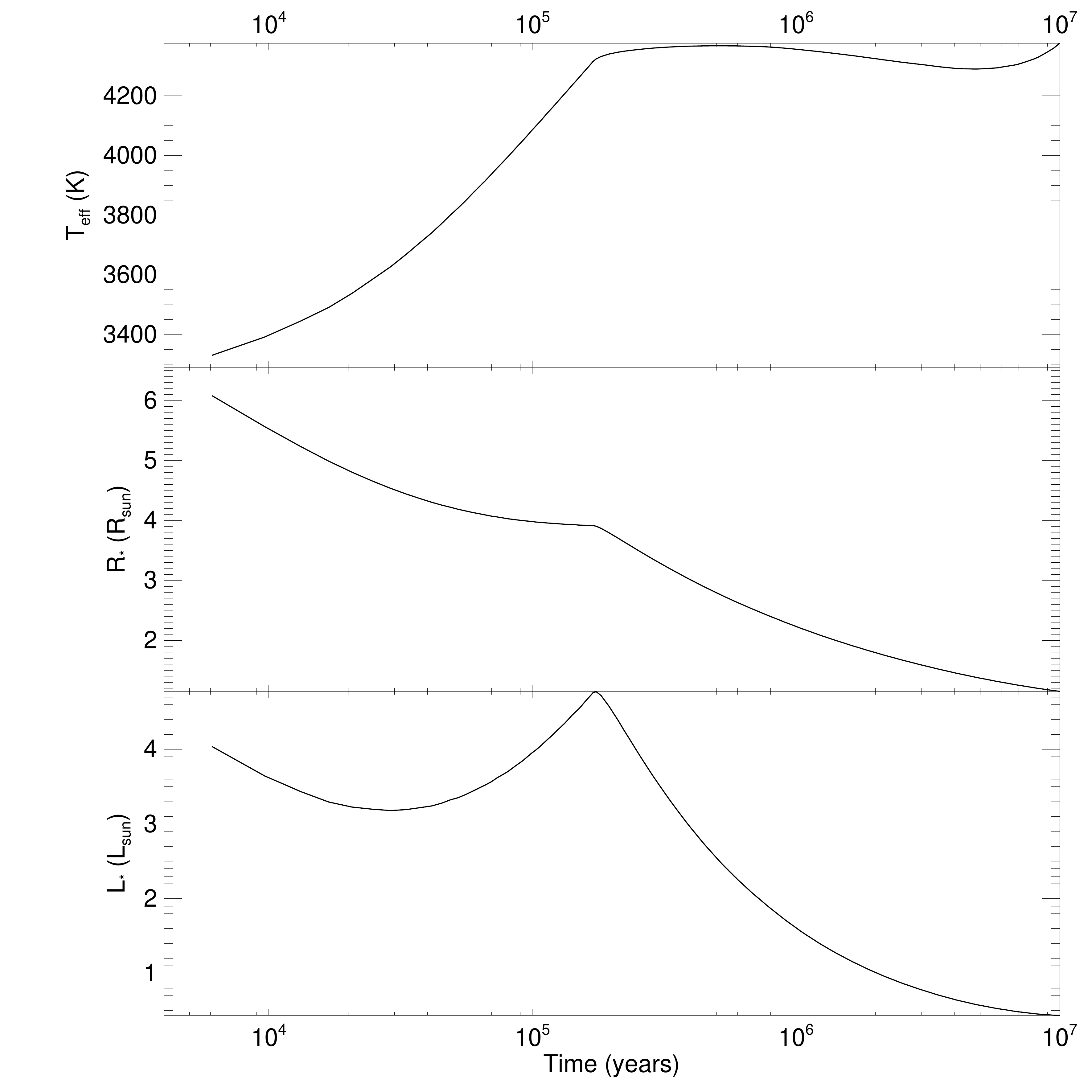}
\end{center}
\caption{Time evolution of the star effective temperature, radius, and luminosity.}
\label{trletoile}
\end{figure}

\begin{figure}[htbp!]
\begin{center}
\includegraphics[width=8cm, clip=true]{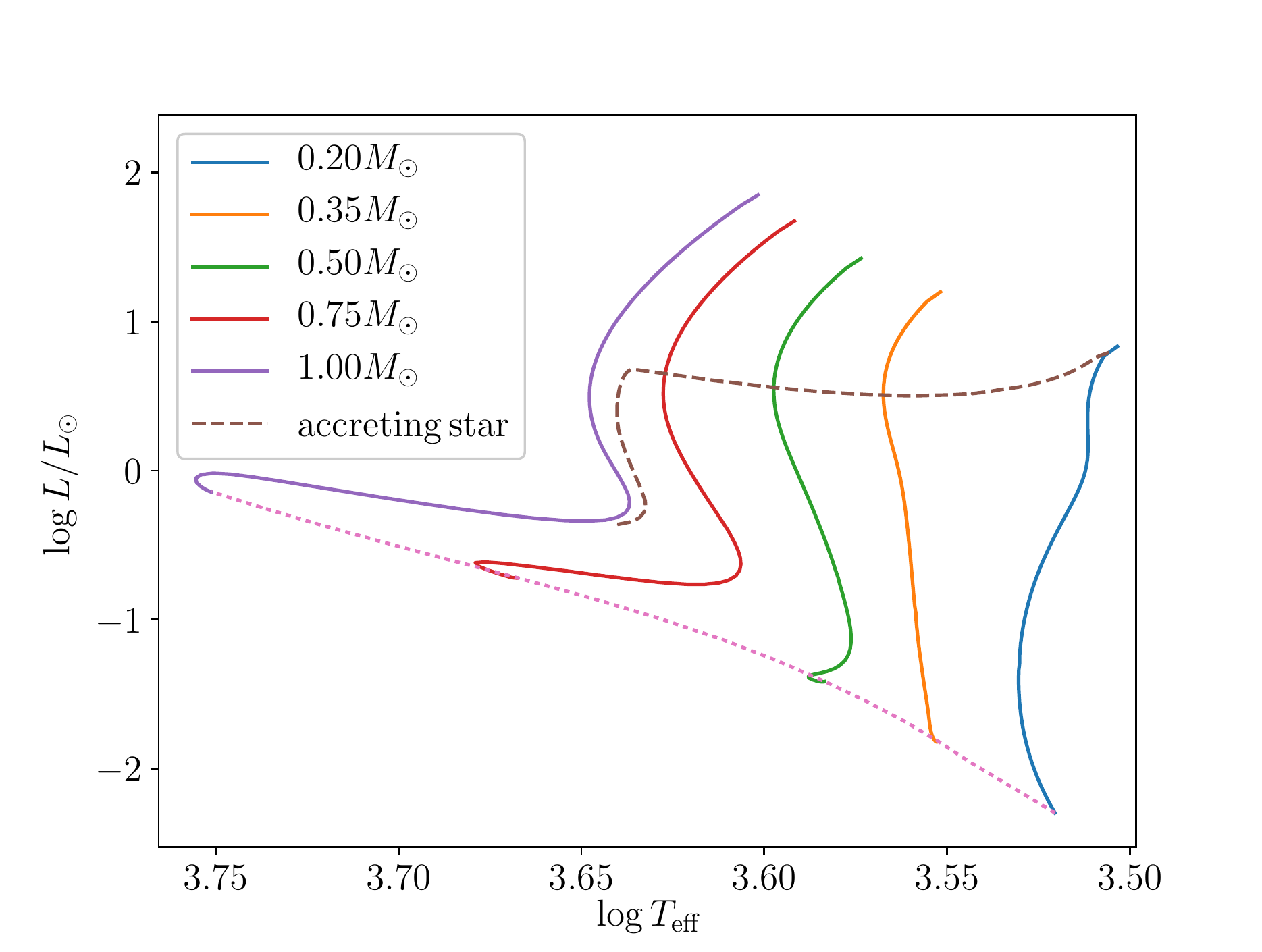}
\end{center}
\caption{Evolutionary track on the HR diagram of the protostar at the center of the disk (brown dashed line). Full lines indicate classical evolutionary tracks at constant mass and the dotted line indicates the ZAMS.}
\label{fig:hr}
\end{figure}

Though we do not directly model the evolution of the star, the interpolation of the star characteristics over pre-calculated stellar evolutions allows us to follow the evolution of its mass (Figure \ref{metoile}), temperature, radius, and luminosity (Figure \ref{trletoile}). The evolutionary track on the HR diagram is shown in Figure \ref{fig:hr}, together with some pre-main sequence tracks at fixed mass. The star approximately follows  the evolution of an accreting protostar with $\dot{M} = 10^{-5} M_{\odot} {\rm yr}^{-1}$ \citep[as seen in, e.g. , ][]{palla90}; however, the luminosity of the star is higher than an accreting protostar at the low-mass end ($M_{*} < 0.4 M_{\odot}$): the modeled star begins higher in the Hayashi track, meaning that its luminosity is closer to that of an accreting protostar later, at higher mass. This difference in luminosity, though not physical, is only temporary; it is just an effect of the initial conditions, in order to have a realistic luminosity in the mass range of interest. We plan to extend this study to the more realistic case where the accreting protostar and the disk are modeled in a self-consistent way.

Starting with an initial star+disk system of total mass $0.2 M_{\odot}$, we reach a total mass of $1 M_{\odot}$ after 170,000 years of evolution. At this date, we set $\dot{M}$ to 0, but note that the star has not yet reached its final mass: it is only $0.84 M_{\odot}$, and the disk mass is $0.16 M_{\odot}$, meaning the disk-to-star mass ratio is around  0.19. We verify that the Toomre instability can only appear transiently in the outer disk around the end of the collapse phase, and that it cannot affect the accretion rate of the disk onto the central star.

In the first phase (the gravitational collapse), the disk and star masses grow linearly. The disk gains mass from its surface while, at its inner edge, it loses some material that falls onto the star. After the cloud has emptied, it can no longer provide material  and the disk can only yield gas and dust to the star by its inner edge: the disk mass slowly decreases while the star tends asymptotically towards $1 M_{\odot}$ (see the bend in the evolutionary track in Figure \ref{fig:hr}). After 1.64 Myr, the disk-to-star mass ratio is equal to 0.087, which corresponds to the one that \citetalias{baillie14}, \citetalias{baillie15} and \citetalias{baillie16} used as their initial disk following the MMSN \citep{weiden77, hayashi81}.

From Figure \ref{metoile} and Equation \ref{rce}, one can directly estimate the evolution of the centrifugal radius which tends to 10.5 AU at the end of the collapse phase. In conditions similar to \citet{yang12} (see their Figure 2), the mass evolution of the central star and disk are consistent with their results. During early evolution (before 10,000 years), when the centrifugal radius is not significantly larger than the stellar radius, the collapsing material falls directly onto the star. This explains why the disk seems to start growing a little later than the star. Subsequently, when the centrifugal radius becomes larger, the disk receives more material and grows until the cloud gas reservoir is empty.

As a consequence of the observed mass accretion rates, the star also evolves: its characteristics vary during the collapse phase before stabilizing at 170,000 years, as can be seen in Figure \ref{trletoile}.

The star temperature increases from 3400 K to 4400 K, and its radius decreases from $6 R_{\odot}$  to $1 R_{\odot}$ after 10 Myr. The luminosity first decreases during the first 30,000 years, before increasing up to $5 \mathcal{L}_{\odot}$ at the end of the collapse phase. After about 2 Myr of evolution, the star is very similar to the one used in \citetalias{baillie14}, \citetalias{baillie15} and \citetalias{baillie16}, though with a slightly lower radius than the $3 R_{\odot}$ taken in these previous works. The final star corresponds to a classical T-Tauri pre-main sequence star, slightly fainter than its luminosity at the zero-age main sequence (ZAMS).

\subsection{Disk evolution}
\label{evol}
Photo-evaporation due to the intense stellar irradiation will dissipate the gas of the disk after a few million years \citet{font04}, \citet{alexander07}, \citet{alexander09} and \citet{owen10}. However, as we do not take that effect into consideration in the scope of the present paper, we are able to pursue our simulations over 10 Myr. Outputs of our code should be considered with great care for the last few million years however, given that the modeled gas may have dissipated at that time.

\citetalias{baillie16} not only modeled the evolution of the surface mass density, but also derived the thermodynamical profiles of such disks: from the pressure scale-height to the surface geometry and the midplane temperature. In particular, they described how the sublimation of the various dust species across the disk generates a temperature plateau at the sublimation temperature of each of these species (corresponding to the drops in opacities as a function of the midplane temperature) and triggers shadowing of the disk surface. As the disk evolves and cools down, these plateaus drift inward; the silicate sublimation region soon being confined at the inner edge of the disk. In the present paper, we refer to the mean locations of these sublimation plateaus as "sublimation lines".

\begin{figure}[htbp!]
\begin{center}
\includegraphics[width=8cm, clip=true]{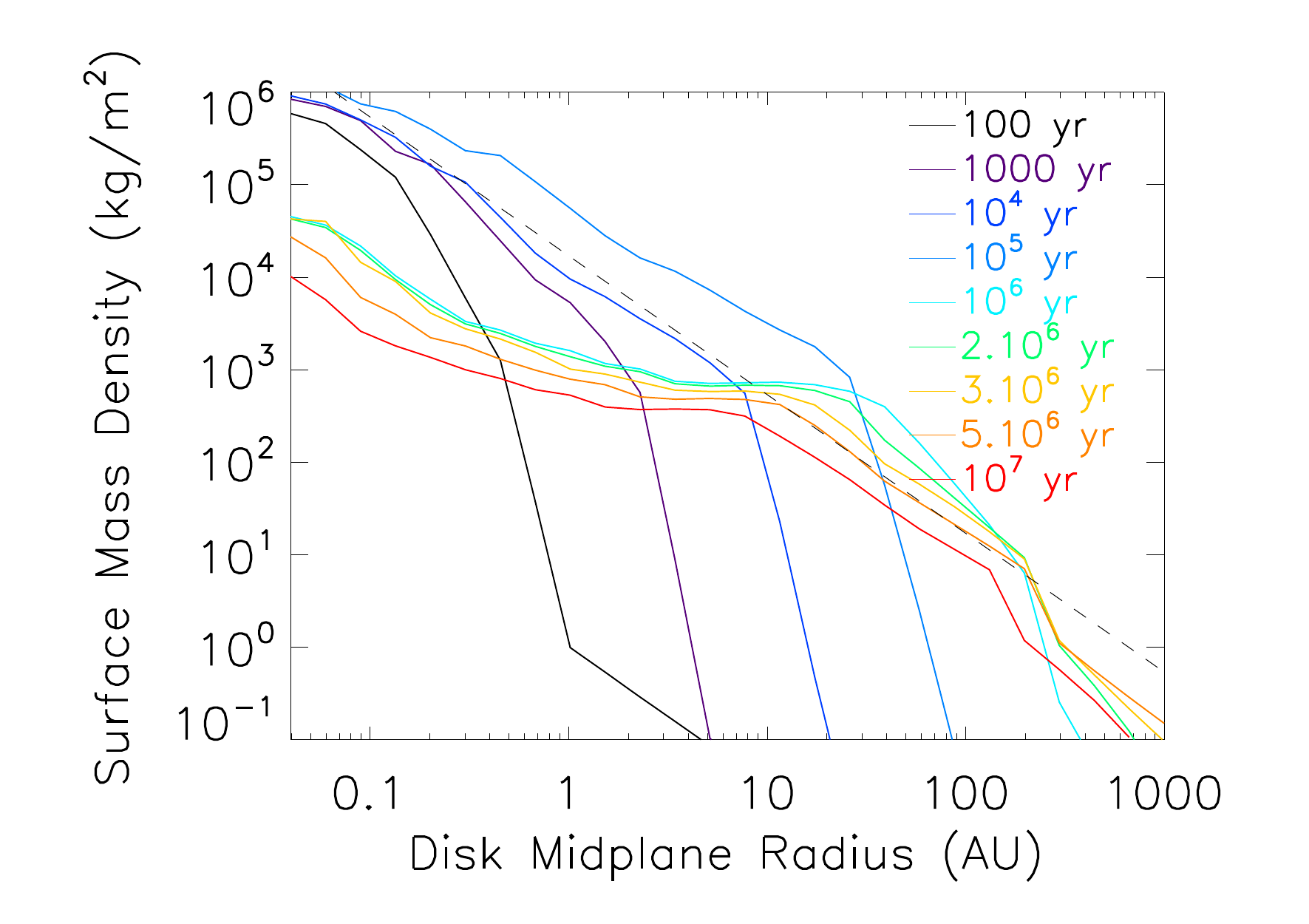}
\end{center}
\caption{Surface mass density radial profile evolution of a disk formed by the collapse of a molecular cloud. The dashed line shows the MMSN defined by \citet{hayashi81}.}
\label{sigt}
\end{figure}

Figure \ref{sigt} shows the evolution of the radial profiles of surface mass density. Early in the disk lifetime, this profile cannot be simply represented by a power law, as was the case for the MMSN disk models. We notice significant changes in the slopes of those profiles until 1 Myr, after which the density profiles become smoother. As the disk gains mass and forms the molecular cloud, its surface mass density grows, especially in the inner disk. The inner viscous heating tends to increase as well as the disk optical thickness, which in turn affects the disk midplane temperature. As a further consequence, the vertical extent of the disk (pressure scale height or photosphere height), and therefore its geometry, are also affected by the infall of cloud material. At the beginning of the simulation (black curve at 100 years), the disk is concentrated inside 1 AU. It then spreads as it gains in mass: after 100,000 years, it reaches several tens of astronomical units with a density at 1 AU, exceeding the initial density of the MMSN disk in \citetalias{baillie15} at the same location by almost an order of magnitude. After 1 Myr (several hundred thousand years after the end of the collapse), the disk evolves viscously in a similar manner to the MMSN: it spreads and its mass decreases since the star is accreting the material that the disk yields at its inner edge. We notice that the asymptotic trend in the surface mass density profile becomes shallower than the MMSN as the disk ages. These decreasing power-law indices are indeed expected since the mass flux becomes increasingly uniform (\citet{lyndenbellpringle74, chiang97, bitsch14} and \citetalias{baillie14}).

\begin{figure}[htbp!]
\begin{center}
\includegraphics[width=8cm, clip=true]{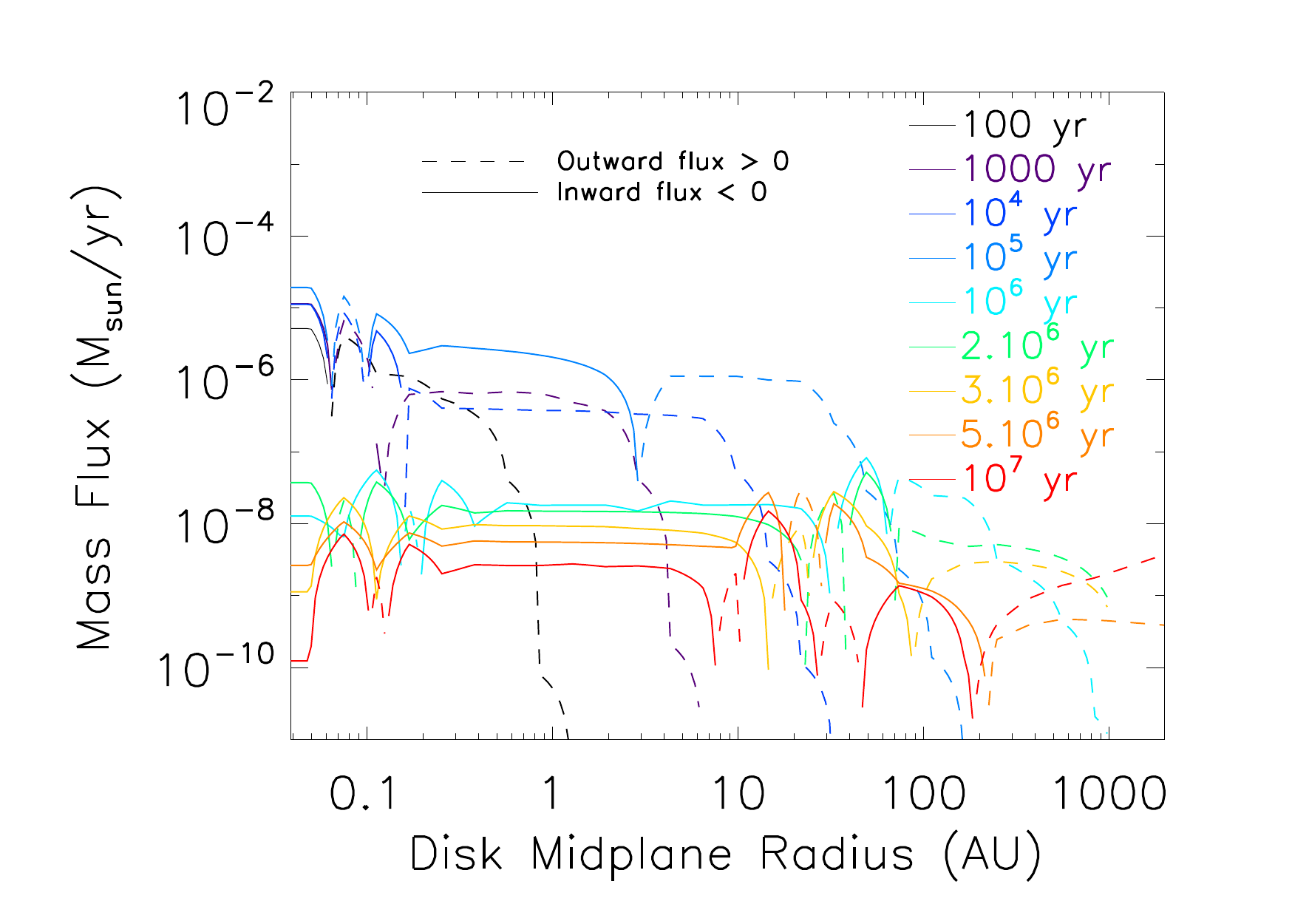}
\end{center}
\caption{Evolution of the viscous mass flux radial profiles for a disk formed by the collapse of a molecular cloud.}
\label{massflux}
\end{figure}

Figure \ref{massflux} indicates a viscous mass accretion rate of approximately a few times $10^{-6} \, M_{\odot}/\mathrm{yr}^{-1}$ after 100,000 years, a few times $10^{-8} \, M_{\odot}/\mathrm{yr}^{-1}$ after 1 Myr, and an order of magnitude lower after 10 Myr. Thus, mass accretion rates are above the estimated ones for the MMSN in the first 100,000 years. After the collapse phase, such a disk has a mass accretion rate similar to that of an MMSN that would be younger by a few million years. In other words, the MMSN corresponds to a phase that occurs a few million years after the start of the collapse.

\begin{figure}[htbp!]
\begin{center}
\includegraphics[width=8cm, clip=true]{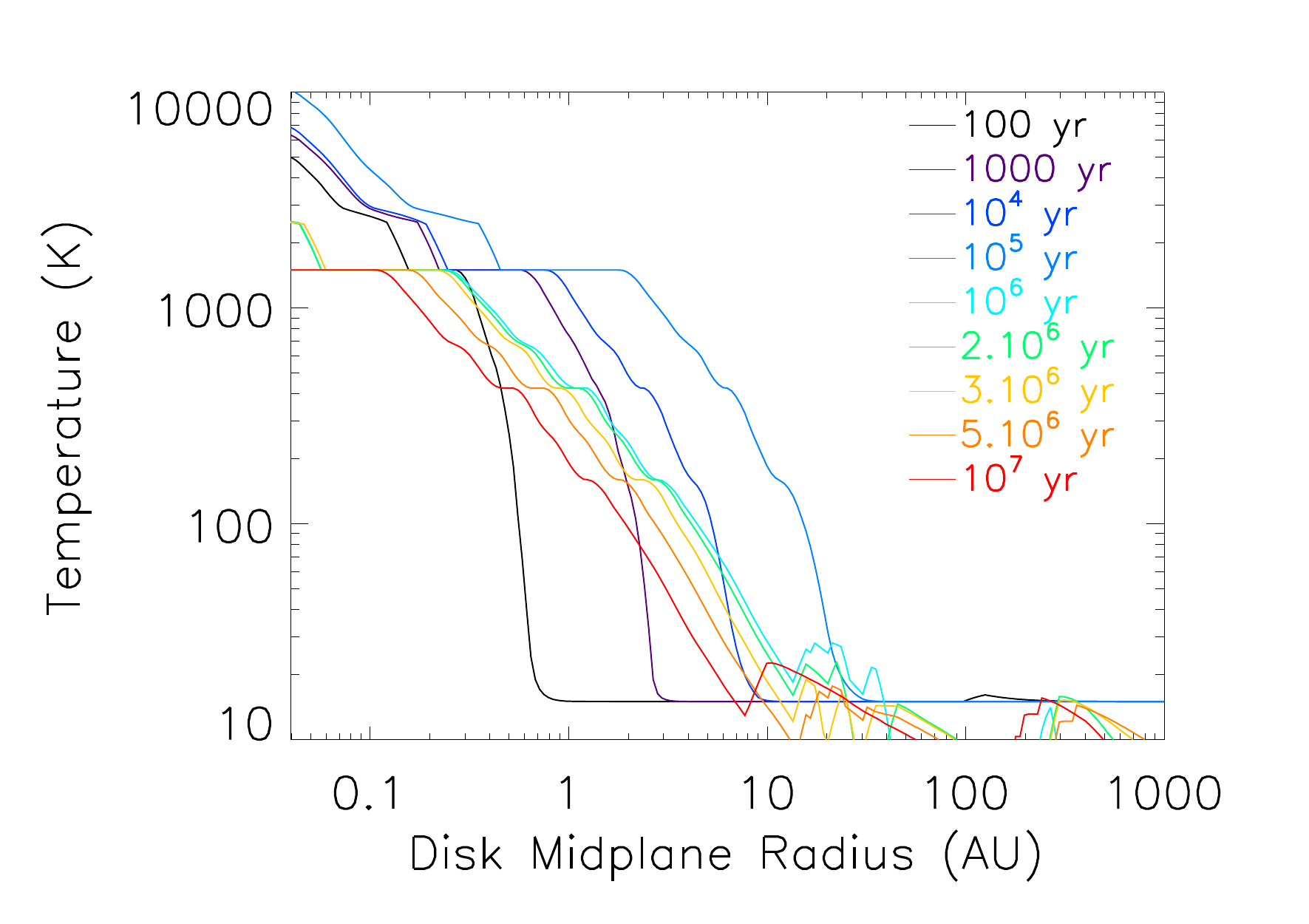}
\end{center}
\caption{Midplane temperature radial profile evolution of a disk formed by the collapse of a molecular cloud.}
\label{tempt}
\end{figure}

Similarly, the evolution of the temperature radial profile (Figure \ref{tempt}) shows that the disk is initially cold and will be heated for a few hundred thousand years while it gains mass from the infall of the molecular cloud. Later, when the collapse phase is finished, the disk follows a similar evolution to that of an MMSN \citepalias{baillie15} and cools. In particular, we notice that the temperature plateaux (where dust species sublimate) drift outward at first and then inward after a few hundred thousand years.

\begin{figure}[htbp!]
\begin{center}
\includegraphics[width=9cm, clip=true]{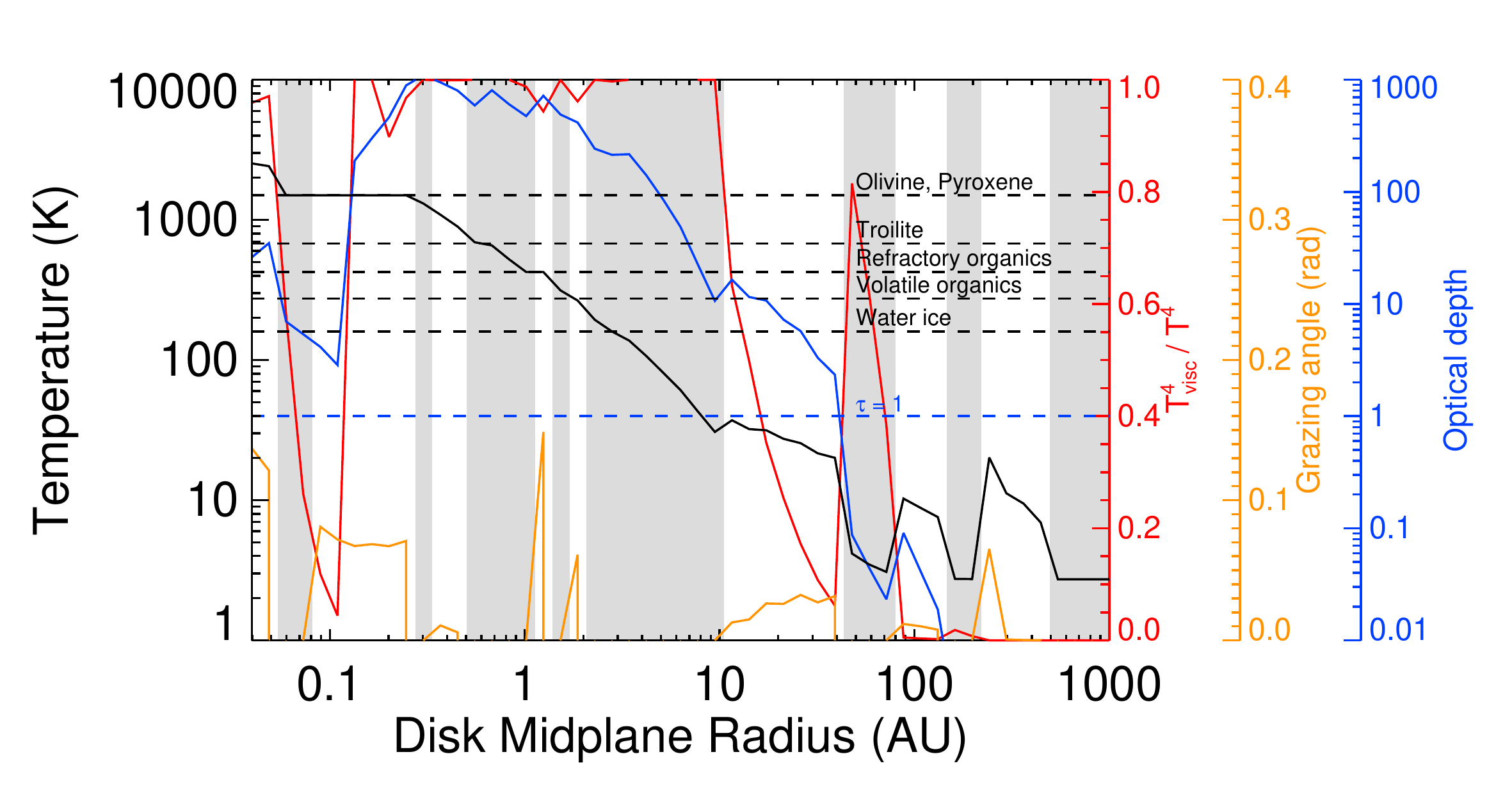}
\end{center}
\caption{Mid-plane temperature radial profile (black) after 1 Myr of evolution with a self-consistently calculated geometry and a full continuous model of opacities. Disk-shadowed regions are displayed in gray. The ratio of the viscous heating contribution to the total heating (viscous heating rate) is presented in red, the grazing angle radial profile in yellow, and the optical depth radial profile in blue.}
\label{profil}
\end{figure}

During the collapse phase, it is worth noticing that the outer regions are the most affected by the cloud envelope radiation term: this imposes a minimal temperature of $T_{\mathrm{cd}} = 15$ K anywhere in the disk and in particular in the outer regions. This results in a lower sensitivity of the outer disk temperature to the variations in the disk optical depth in the regions where the disk is the thinnest.

Given that the density and temperature structures are implicitly related, it is very difficult to estimate a precise error bar on these profiles. However, as \citetalias{baillie15} showed by studying the impact of the radial resolution, density and temperature structures are still sustained despite resolution changes. Therefore, the sublimation zones and planet traps can only marginally be affected for a given set of initial parameters.

Figure \ref{profil} details the various heating contributions after 1 Myr. Irradiation heating locally dominates between 0.1 and 0.15 AU before viscous heating takes over when the disk gains optical depth. This defines an inner heat transition barrier, located slightly below 0.15 AU after 1 Myr. This coincides with an immediately outer shadow region located around 0.3 AU, shortly after the silicate sublimation plateau. The part of the disk inner to 0.1 AU appears to be consistent with the inner disk structures described by \citet{flock16} using radiation hydrodynamical simulations: they expect an inner rim around the silicate sublimation front that would cast a shadow upon the few tenths of astronomical units immediately further out. \citet{flock17} confirmed this in 3D radiation nonideal magnetohydrodynamical simulations. Though these latter studies treat MHD more thoroughly than we do here (they rely on the results of MHD magnetorotational turbulence models while we only consider it implicitly in the turbulent viscosity parameter), the inner disk regions present consistent geometric and thermic structures between MHD and purely hydrodynamical simulations. In addition, the grazing angle radial profile shows that the disk is only marginally irradiated below 10 AU at 1 Myr: the grazing angle hardly passes 0.1 rad. Around 10 AU, the viscous heating rate drops and the grazing angle grows to heat up the disk by a few Kelvins.

Figure \ref{profil} shows a succession of shadowed regions between 0.3 and 10 AU, which starts at the outer edge of the silicate sublimation plateau, similarly to what was observed in \citetalias{baillie15}. Drops in dust-to-gas ratio are consequently expected at the temperature plateaus as dust elements sublimate species by species according to their sublimation temperature. Consistently with our opacity model, all the dust is sublimated for temperatures above 1500 K, leading to a dust-to-gas ratio of zero, whereas for temperatures below 160 K, all the dust species are solid and the dust-to-gas ratio reaches a maximum value of 1\%.

In addition, after 1 Myr, we notice that the temperature structures are shifted outward: the outer edge of the plateau related to the sublimation of the silicates is now around 0.25 AU in the present simulation while it was located around 0.15 AU in the MMSN case. Likewise, the outer edge of the water ice line is now around 3 AU instead of 2 AU in the MMSN case. Nonetheless, the outer heat transition barrier (coinciding with the outer edge of the outermost shadowed region) is still found around 10 AU.

\begin{figure}[htbp!]
\begin{center} $
\begin{array}{c}
\includegraphics[width=8cm, clip=true]{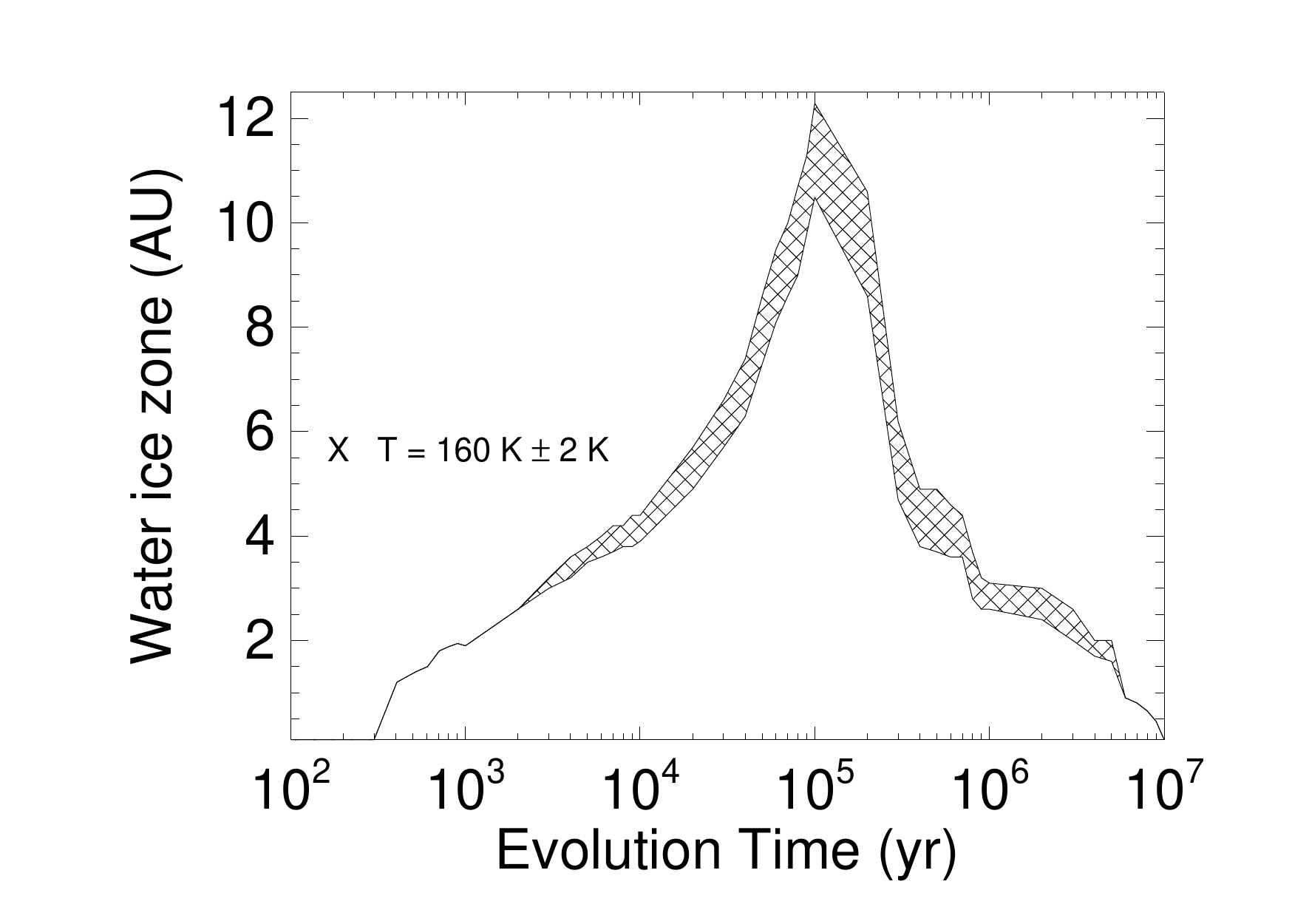}\\
\includegraphics[width=8cm, clip=true]{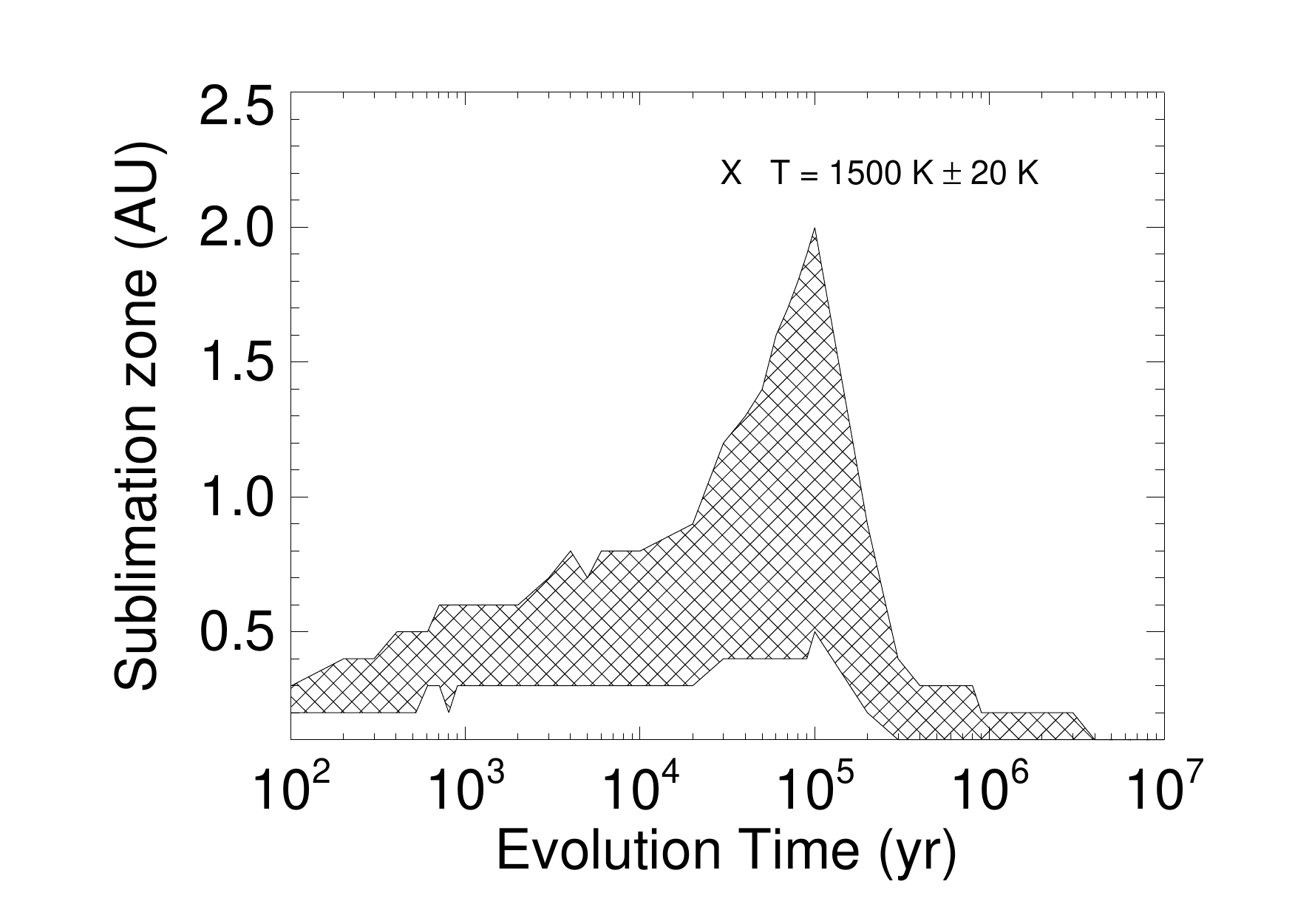}
\end{array} $
\end{center}
\caption{Upper panel: Time evolution of the water ice sublimation region (mid-plane radial location for which the temperature coincides with the water-ice condensation temperature $160 \, \pm \, 2$ K). Lower panel: Time evolution of the silicates sublimation zone (mid-plane radial location for which the temperature coincides with the silicate sublimation temperature $1500 \, \pm \, 20$ K).}
\label{lines}
\end{figure}

We may now follow the evolution of the positions of the sublimation plateaus: the water ice line at $160 \pm 2$ K (Figure \ref{lines} - upper panel) starts initially below 1 AU and moves up to 12 AU at the end of the collapse phase before drifting inward below 2 AU after 3 Myr. Apart from at the beginning of the collapse phase (the first few thousand years), the water ice line is always further away from the star than it was in the MMSN simulations. In addition, its width regularly exceeds 1 AU. The observation is similar for the silicate sublimation line at ($1500 \pm 20$ K) that can be located around 2 AU at the end of the infall before tending towards 0.1 AU after 2 Myr.

\subsection{Equivalent timeline}
\label{timeline}
\begin{figure}[htbp!]
\begin{center}
\includegraphics[width=8cm, clip=true]{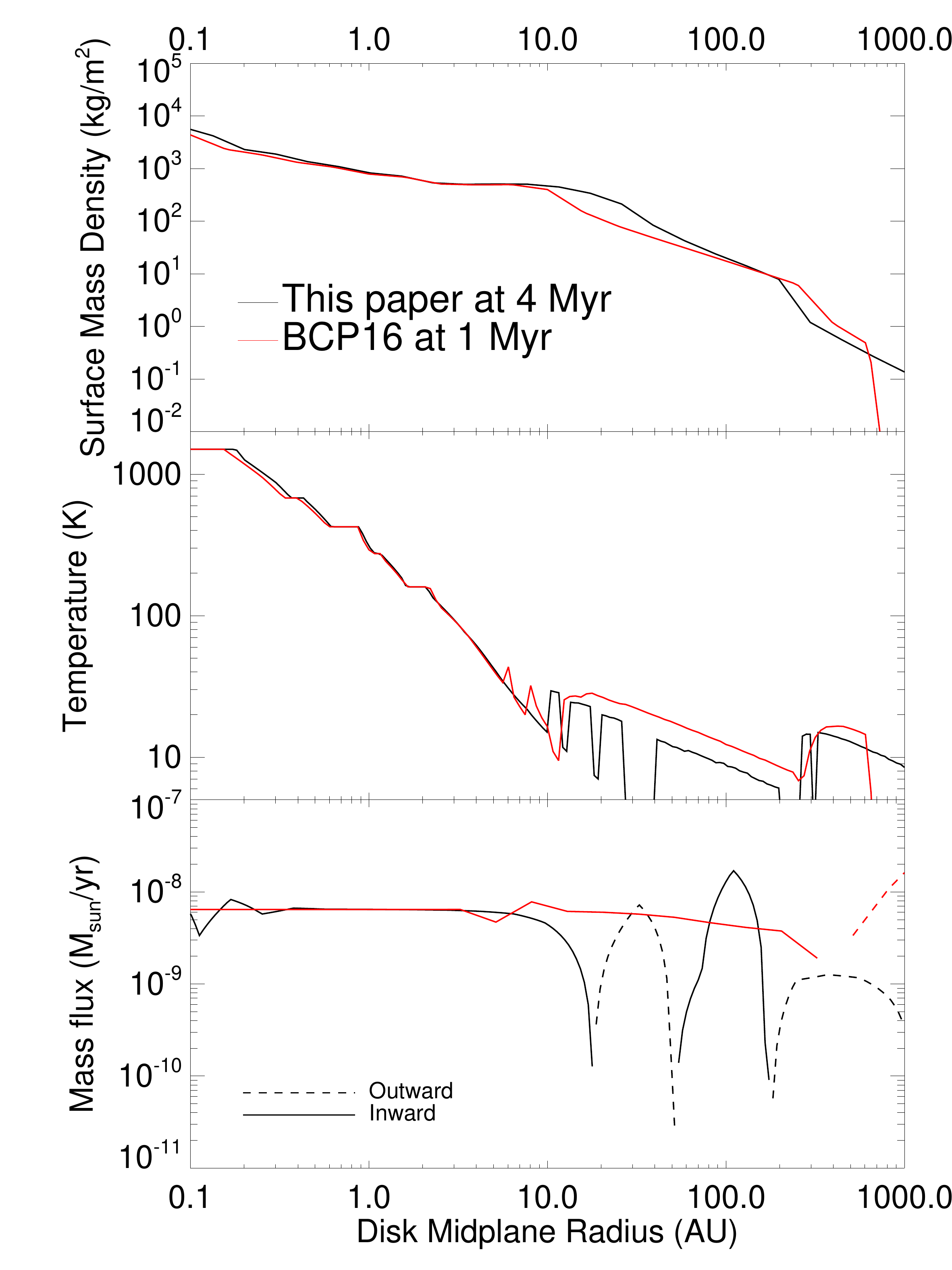}
\end{center}
\caption{Comparison of the disk radial profiles (surface mass density - upper panel, temperature - middle panel, and mass flux - lower panel) for the disk at 4 Myr in the present paper (black line) and a 1 million-year-old disk evolved from an MMSN in \citetalias{baillie16} (red). We notice that the new treatment of the disk self-shadowing induces the presence of temperature drops of up to 10 K in the region 10 - 100 AU that possibly generate narrow heat transition barriers prone to trap planets.}
\label{compbcp16}
\end{figure}

Due to the difference in the initial conditions between the present simulations and those from \citetalias{baillie16}, we have to consider that the initial ages of the disks are not equivalent in these two simulations. Admitting that the disk age can be correlated to its mass accretion rate onto its central star, we should compare disks evolved from different initial conditions at equivalent mass accretion rates rather than elapsed time in the simulation. To that end, we focus on the mass flux profiles across the disks presented in Figure \ref{massflux} of the present paper and Figure 4 of \citetalias{baillie15}. This suggests that a disk evolved from the collapse of the molecular cloud for 500,000 years has a mass flux (and therefore a similar age) that is  similar to that of a disk evolved from an MMSN for 10,000 years. Similarly, a collapse-formed disk after 3 Myr seems equivalent in mass flux to an MMSN disk after 100,000 years; a collapse-formed disk after 4 Myr would be equivalent to an MMSN disk after 1 Myr as shown in Figure \ref{compbcp16}; and a collapse-formed disk after 7 Myr would be equivalent to an MMSN disk after 5 Myr. The density and temperature profiles (Figures \ref{sigt} and \ref{tempt} of the present simulation vs. Figures 3 and 5 of \citetalias{baillie15}) tend to validate these associations, along with the evolutions of the sublimation lines (Figure \ref{lines} of this paper vs. Figures 11-12 of \citetalias{baillie15}). This suggests that a disk evolved from an MMSN could coincide with a collapse-formed disk that would be older by at least 2 Myr.

In the present paper simulations, it takes a few hundred thousand years for the disk to reach a profile similar in shape to the ones evolved from an already formed MMSN. However, comparing this disk with an early MMSN-like disk is not as pertinent as comparing disks that are close to reaching their steady state. Such a comparison shows that the MMSN simulation seems to follow the collapse simulation by a couple of million years.

\subsection{Type-I migration}
\label{torque}
A planet excites resonances (\citet{goldreich79,ward88,arty93,jc05}) across the disk: Lindblad resonances caused by the action of the spiral arms induced by the planet, and corotation resonances due to the horseshoe region around the planet. Discounting the back-reaction of the planet on the disk structure, we can derive the resonant torques that the planet exerts on the disk and then calculate the reaction torque exerted by the disk on the planet. Using the expressions derived in Appendices \ref{torquelin} and \ref{torquecor} for the Lindblad and corotation torques, we derive the total torque that a hypothetical planet of mass $M_{\mathrm{P}}$, located at a radial distance $r_{\mathrm{P}}$ from the central star, would receive from a viscously evolving disk.

\begin{equation}
\label{gammatot}
\Gamma_{\mathrm{tot}} = \Gamma_{\mathrm{Lindblad}} + \Gamma_{\mathrm{corotation}}
.\end{equation}

Our model assumes a constant and uniform turbulent viscosity $\alpha$, and does not consider any deadzone or cavity inside the disk: we may neglect variations in the viscosity and apply the torque formulas detailed in Appendices A and B. In addition, these torque expressions are consistent with the ones used in the previous works we compare to.

Since the Lindblad torque is in principle dominated by the variations of the temperature compared to the ones of the density, a disk for which the density and temperature profiles are decreasing with the distance to the star will experience a negative Lindblad torque, leading to a possible inward migration. However, the corotation torque may counter the Lindblad torque and slow down or reverse the migration. In the temperature plateaus, the temperature variation is much lower than the density variation, which could lead to a simplification of the expressions provided in the Appendix. However, the strong influence of the cut-off and saturation effects makes it difficult to predict the final sign of the total torque. The edges of the regions where the migration is outward define locations where planetary embryos converge (the so-called "planet traps"), and places from which embryos run away ("planet deserts"). \citetalias{baillie15} showed that such planet traps usually appear at the outer edge of the sublimation regions of the different dust elements, or at the heat transition barrier between a region dominated by viscous heating and a region dominated by irradiation heating. This barrier coincides with the outer edge of the most external shadowed region (region for which the photosphere is no longer directly irradiated by the star).

The planet mass is another key parameter for the estimation of torque amplitude: the normalization coefficient $\Gamma_{0}(r_{\mathrm{P}})$ depends on the planet mass through Equation \ref{gamma0}, and furthermore the saturation of the corotation torques is a function of the planet mass. This second effect is the only one due to the planet that is visible in the migration maps of Section \ref{mmaps} which are normalized by $\Gamma_{0}(r_{\mathrm{P}})$. We note that the evolving temperature and density of the disk also affect these maps as the disk ages. \citetalias{baillie16} stated that low-mass planets ($< 10 M_{\mathrm{\oplus}}$) cannot have a positive corotation torque large enough to balance the negative Lindblad torque near the sublimation lines, and therefore remain in inward migration. For more massive planets located near the sublimation lines, the positive corotation torque increases, and induces a positive total torque. Further out in the disk, near the heat transition barrier, a locally positive temperature gradient induces an outward migration for planets more massive than a few tens of Earth masses. The cut-off at high viscosity introduced by \citet{paar11} (Equations \ref{GG} and \ref{KK}) reflects a drop of the corotation torque for massive planets, as illustrated by the results of \citetalias{baillie16} showing that planet traps appear to be correlated with sublimation lines only for planet masses between 10 and 100 $M_{\mathrm{\oplus}}$. However, additional traps related to the heat transitions can be found for planets not in this range.

\subsection{Type-II migration}
If the planet is massive enough and/or the disk aspect ratio $h/r$ is low enough, the planet may open a gap in the gas of the protoplanetary disk. \citet{crida06} derived an empirical criterion for the gap-opening condition:

\begin{equation}
\label{eq15crida06}
\frac{3 h_{pr}}{4 R_{Hill}} + \frac{50}{q \mathcal{R}} \lesssim 1
,\end{equation}

with $R_{Hill} = r_{\mathrm{P}} \left( \frac{M_{\mathrm{P}}}{M_{*}}\right)^{1/3}$ , which is the Hill radius, $\mathcal{R} = \frac{r_{\mathrm{P}}^{2} \Omega_{\mathrm{P}}}{\nu_{\mathrm{P}}}$ , the Reynolds number, and $q = \frac{M_{planet}}{M_{*}}$, which is the mass ratio of the planet to the star.

\citetalias{baillie16} found that in the case of the viscous evolution of an MMSN, gaps may be opened at the places where the dominant heating process switches from viscous heating to irradiation heating: for instance, this concerned planets more massive than $170 M_{\mathrm{\oplus}}$ around 15 AU in the early disk (10,000 years), or planets above $180 M_{\mathrm{\oplus}}$ located around 7 or 9 AU after 1 Myr. This could be explained by the trough in aspect ratio that \citetalias{baillie15} observed in their Figure 6.

In the following sections, we study how planet traps and deserts, and also gas gaps, are affected by the disk origin (molecular cloud collapse) and how it may impact the growth scenarios of planetary embryos.

\subsection{Planetary trap evolution}

\begin{figure}[htbp!]
\begin{center}
\includegraphics[width=7cm, clip=true]{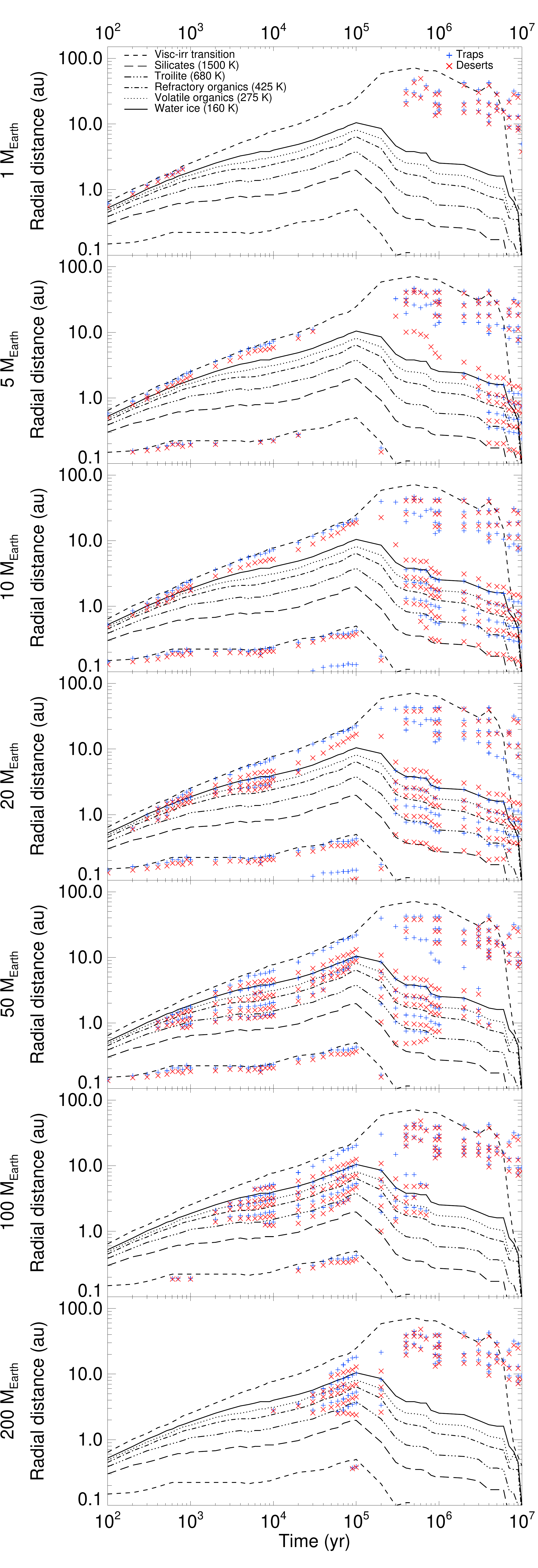}
\end{center}
\caption{Time evolution of the migration trap (blue "+") and desert (red "x") positions. The sublimation line positions and the heat transition radius are represented with the black dotted and dashed lines. Each subfigure shows the traps and deserts for a given planet mass.}
\label{traptime}
\end{figure}

Based on the simulated disk profiles, we derive the torques that hypothetical planets of given masses and radial distances to the star would experience, and we extract the planet trap and planet desert positions for each of the given times. Focusing on the planetary formation region, Figure \ref{traptime} details the locations of these traps and deserts for planet masses between 1 and 200 $M_{\mathrm{\oplus}}$.

As described in \citetalias{baillie16}, the radial profile of the total torque at a given time shows a background of inward migration with possibly a few intervals of outward migration. Therefore, we expect each desert to be accompanied by a trap slightly further out, as the total torque returns to a background negative value outside of this outward migration range.

A low-mass planet ($1 M_{\mathrm{\oplus}}$) can only be trapped at the outer heat transition frontier, either during the first 1000 years or after the end of the collapse. In the second case, such a small planetary embryo would be trapped at distances from the star of between 8 and 40 AU.

For planets of $5 M_{\mathrm{\oplus}}$ or more, we notice a second recurrent population of desert+trap located at the inner heat transition frontier around 0.2 AU that appears very early in the simulation (during the collapse phase). Thus, a planetary embryo with the mass of a super-Earth could get trapped on the orbits of Mercury or Venus, in the very early instants of a disk formed by gravitational collapse. In addition, new traps begin to appear at the sublimation lines of the dust species after 2 Myr.

More massive planets ($10 M_{\mathrm{\oplus}}$) present traps at the same sublimation lines as previously noted, though they appear earlier, after 400,000 years. Trapping planets at the sublimation lines may happen earlier for more massive planets. Planets of $100 M_{\mathrm{\oplus}}$  may even be trapped at all the sublimation lines, even during the collapse phase. We also note that the inner heat transition frontier can only trap embryos of up to $100 M_{\mathrm{\oplus}}$.

Planets more massive than $200 M_{\mathrm{\oplus}}$ can no longer be trapped below 1 AU: the outer heat transition barrier remains the only possibility for these planets to be trapped after the collapse phase.

Comparing Figure \ref{traptime} with Figure 2 from \citetalias{baillie16}, we notice that traps now appear later than they used to in the case of the MMSN evolution. However, they seem to survive longer: for instance, a planet of $10 M_{\mathrm{\oplus}}$ can be trapped after 400,000 years while it was possible as early as 20,000 years in the MMSN disk simulations. Similarly, planets of $100 M_{\mathrm{\oplus}}$ could remain trapped at the sublimation lines until 10,000 years in an MMSN disk while they can now survive the first million years in the case of a collapse-formed disk. This is consistent with the time delay observed in Section \ref{timeline} between the timelines of the two simulations. Nevertheless, the present simulations provide indications of the trapping possibilities for planetary embryos during the phase when the collapse-formed disk cannot be modeled by a stage of the evolution of an MMSN. Thus, an early trapping at the inner heat transition frontier enables the transient survival of more massive planets until this trap eventually falls down onto the central star at the end of the disk simulation. Finally, both simulations agree on the possibility of trapping (very) massive planets at the outer heat transition barrier.

\subsection{Migration maps}
\label{mmaps}
\begin{figure}[htbp!]
\begin{center}
\includegraphics[width=8cm, clip=true]{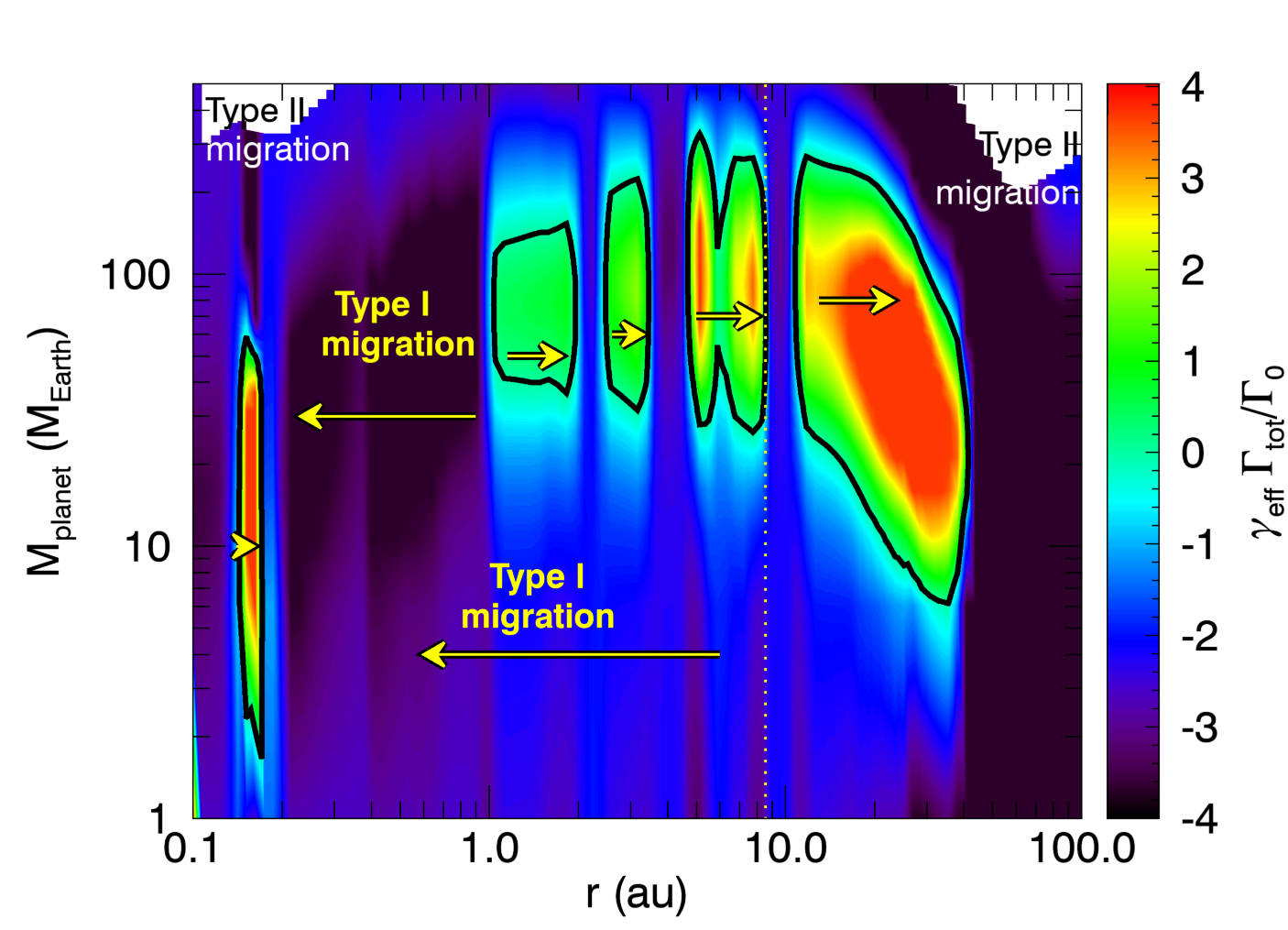}
\end{center}
\caption{Migration torque of a protoplanet with given radial distance to the central star $r_{\mathrm{P}}$ and mass $M_{\mathrm{P}}$, in a protoplanetary disk after 200,000 years of evolution. Black contours (zero-torque contour) delimit the outward migration conditions while the rest of the migration map shows inward migration. Planetary traps are located at the outer edges of the black contours while planetary deserts are at their inner edges. The yellow dotted line marks the water ice line and the white area verifies the criterion from Equation \ref{eq15crida06}.}
\label{mmap200000}
\end{figure}

\begin{figure}[htbp!]
\begin{center}
\includegraphics[width=8cm, clip=true]{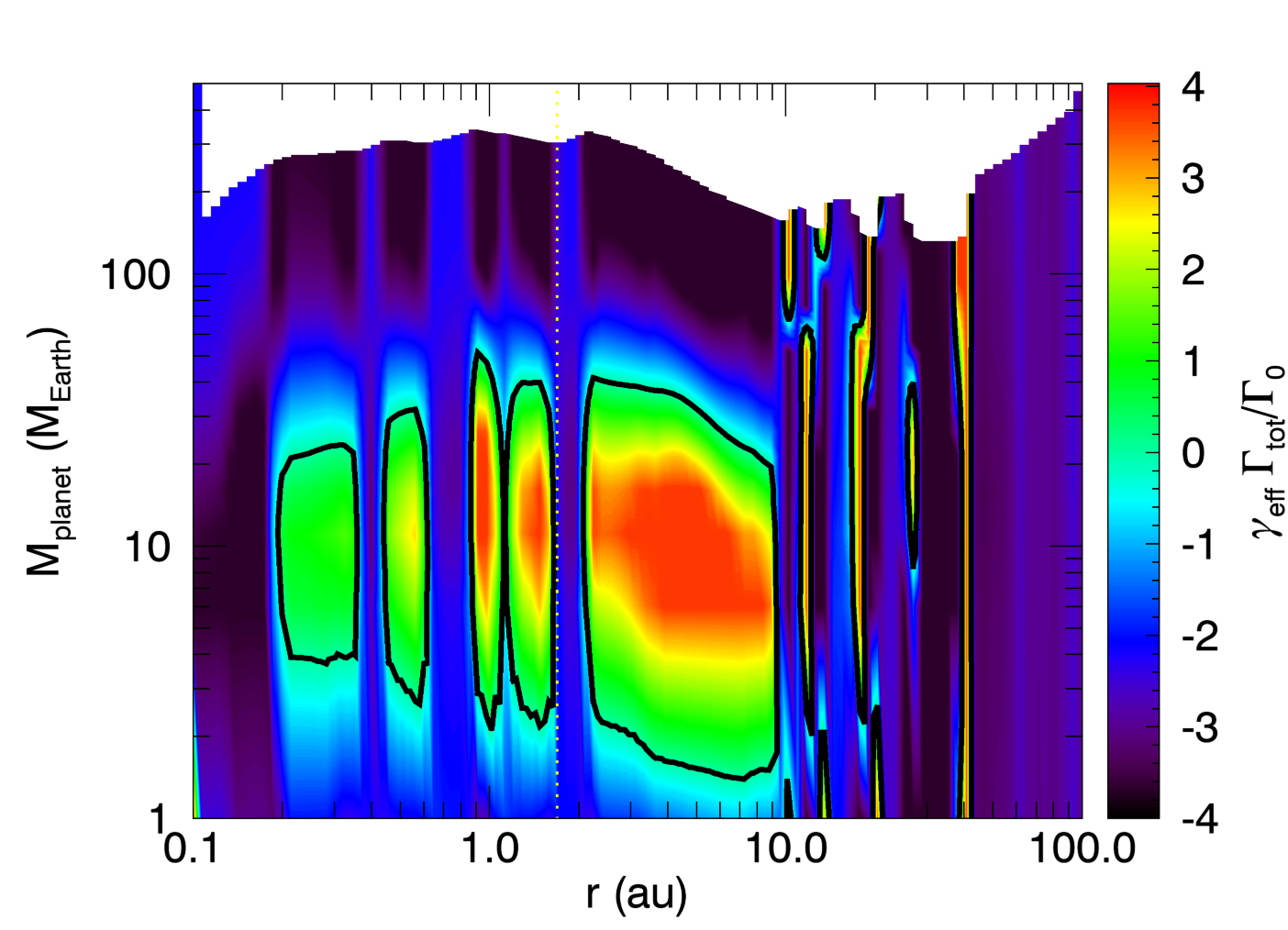}
\end{center}
\caption{Migration map after 4 Myr of evolution. The legend is the same as in Fig. \ref{mmap4000000}.}
\label{mmap4000000}
\end{figure}

In the present section, we present migration maps (Figures \ref{mmap200000}-\ref{mmap4000000}) that display the normalized total torque exerted by the disk on a putative planet as a function of its radial distance $r_{\mathrm{P}}$ to the central star and its mass $M_{\mathrm{P}}$. Torques are normalized by $\Gamma_{0}$ (Equation \ref{gamma0}). In this distance-mass representation, the blue background stands for a negative total torque synonymous of an inward migration of the planet. Red zones on the contrary show closed regions of outward migration. The contours of these regions are the zero-torque radii: inner edges of those zones are the planet deserts while outer edges are the planet traps. Figure \ref{mmap200000} summarizes this by indicating the direction of migration with yellow arrows. In addition, the white area is the region where Equation \ref{eq15crida06} is verified, that is, where a planet is massive enough to open a gap and enter type-II migration. Finally, the water ice sublimation line is symbolized as a yellow dotted line.

As detailed in \citetalias{baillie16}, a trapped planet gaining mass remains trapped at least until it reaches the upper mass limit of that trap. Beyond that limit, an embryo would resume migrating inward until it reaches an inner trap or falls onto the star.

Figure \ref{mmap200000} shows different possibilities for trapping planetary embryos after 200,000 years. First, we notice three traps located at the icelines: the troilite trap is located at 1.9 AU for $38 \,M_{\mathrm{\oplus}} \, < \, M_{\mathrm{P}} \, < \, 145 \, M_{\mathrm{\oplus}}$ and the refractory organics trap is at 3.4 AU for $35 \, M_{\mathrm{\oplus}}\, < \, M_{\mathrm{P}} \, < \, 200 \, M_{\mathrm{\oplus}}$. Outward migration zones due to the volatile organics and water ice lines merge to allow planet trapping between 5 and 8.5 AU for planets between 30 and more than 250 $M_{\mathrm{\oplus}}$. Another trap appears to be located further away, where the irradiation heating becomes dominant (11-45 AU). A very narrow fifth zone of outward migration can be spotted with a trapping possibility slightly below 0.2 AU for planets between 2 and 60 $M_{\mathrm{\oplus}}$, due to the inner heat transition frontier as can be seen in Figure \ref{traptime}. In addition, planets more massive than 300 $M_{\mathrm{\oplus}}$ (though very unlikely to exist at this stage) may open a gap below 0.3 AU, as well as planets more massive than 200 $M_{\mathrm{\oplus}}$ between 40  and at least 100 AU. These gaps cannot be accessed by a trapped planet as was the case for most of the type-II migration regions obtained by \citetalias{baillie16} with the MMSN disk model. However, \citet{crida17} suggests that planets more massive than a few tens of Earth masses may undergo a runaway growth that may help these planets to reach the gap-opening area fast enough to avoid falling onto the star.

We also notice that our migration maps present more regions of outward migration than the work of \citet{bitsch15a} for instance. This can be explained by the fact that these regions are correlated with the sublimation lines of dust elements and that our opacity model involves more sublimation transitions than the \citet{bell94} model they used in their work. In addition, the absence of heat diffusion in our code enhances the temperature gradients, consequently helping to build traps.

Figure \ref{mmap4000000} shows the traps after 4 Myr, that is, more than 3.8 Myr after the end of the collapse phase. Two groups of traps can be identified. The ones that are inner to the water ice line (yellow dashed line) are associated with some of the sublimation lines, while those that are outer (around 9.5 AU for planets of a few tens of Earth masses, and between 10 and 40 AU) are associated with the heat transitions and the subsequent exit of the disk self-shadow, as seen in Figure \ref{profil}. Traps at the volatile organics line (1.2 AU) and water ice line (1.7 AU) are only efficient on planets less massive than 47 $M_{\mathrm{\oplus}}$. At 0.65 AU (refractory organics line), only planets between 3 and 30 $M_{\mathrm{\oplus}}$ can be trapped, while at 0.35 AU (troilite line), planets of 4 to 25 $M_{\mathrm{\oplus}}$ may be saved. Moreover, it is possible at this date for a planet to open a gap and enter type-II migration anywhere assuming it is massive enough. This figure can be directly compared to Figure 5 of \citetalias{baillie16}  (MMSN after 1 Myr), according to our derivation of the equivalent timeline (Section \ref{timeline}).

%%%%%%%%%%%%%%%%%%%%%%%%%%%%%%%%%%%
%4444444444444444444444444444444444
\section{Discussion} \label{disc}

\subsection{Protoplanetary disk formation}

Most of the previous works that studied the numerical evolution of protoplanetary disks would initiate their numerical simulations with a disk already formed and generally following the MMSN model. This model can be criticized by questioning the positions of the planets at the time of their formation compared to their present positions \citep{crida09}: here, we address this discrepancy by considering the initial parameters of the molecular cloud: its temperature, angular velocity, and mass. 

An extensive analysis of the infall parameters (total mass, mass accretion rate, temperature of the cloud, and turbulent viscosity of the disk) could be the object of a full follow-up study. However, even though the locations of the temperature plateaus and shadow regions are likely to be shifted by
varying the initial conditions, these structures will still appear in the vicinity of the sublimation lines of the major dust components in the disk. In addition, as mentioned in \citet{baillie14iaus}, an increase in $\alpha_{\mathrm{visc}}$ will mainly accelerate the disk viscous evolution while it may allow planetary embryos to desaturate their corotation torque for larger masses, therefore allowing more massive planets to become trapped in more turbulent disks.

In addition, we consider now a joint evolution model for the disk and the star. Thus, the collapse phase lasts for 170,000 years before the disk begins to spread viscously, similarly to the MMSN simulations. However, we retrieve the observational asymptotic trend of the MMSN after the first million years. Though the date at which the collapse ends is not affected by the approximation that neglects the accretion luminosity of the star, the stellar growth could be underestimated due to the lower temperature of the disk that slows down its viscous spreading in the absence of accretion luminosity. We estimate that the MMSN is similar to a disk that would have formed by collapse of a molecular cloud between 2 and 3 Myr ago. Therefore, the first millions of years of evolution after formation of a protoplanetary disk are determined by the collapse and cannot be modeled starting from an MMSN. However, 2 Myr after the beginning of the collapse, the later disk evolution could be approximated starting from an MMSN.

\subsection{Favorable conditions for CAI formation}

As we see in Section \ref{res}, the inner disk is now hotter, over a longer period. The midplane temperature at 0.2 AU now exceeds the silicate sublimation temperature for at least the time of the collapse phase. Moreover, at the same time, the mass flux is directed outward in the inner part of the disk (below 10 AU) until the end of the collapse phase. In the MMSN simulations, these conditions could only be found in the very inner regions of the disk during the first 10,000 years.

In the case of a disk built from collapsing material from the molecular cloud, \citet{yang12} showed that refractory materials such as calcium-aluminum-rich inclusions (CAIs) could form during the collapse phase, with a maximum around the end of the infall when the sublimation lines are the furthest away from the star. They would then be transported further out in the disk where they could be preserved in primitive bodies. In line with the model of this latter study, the disk structures presented here   validate the two specific conditions for this process to happen: reaching the dust sublimation temperature in the inner disk and being able to transport the formed materials outwards, towards colder regions where they can recondense. This must happen on longer timescales and over wider regions of the disk than in the MMSN case. This should certainly provide favorable conditions for the models of CAI formation described in \citet{charnoz15}.

\subsection{Planet migration and survival}
\label{savedplanets}

Migration maps (Figures \ref{mmap200000}-\ref{mmap4000000}) show that after the collapse phase, any planet more massive than $2 M_{\mathrm{\oplus}}$ will encounter a planet trap during its inward migration. After 4 Myr, the embryos that may fall onto the star are the ones that would not have reached $2 M_{\mathrm{\oplus}}$ or that would have grown to over a few tens of Earth masses very rapidly. Planet growth is one of the critical process that may force a planet out of the traps.

Based on their locations, traps naturally select which planets they allow to escape type-I inward spiraling, with respect to their mass: different types of planets may be saved.

\begin{itemize}
\item The inner heat transition barrier may temporarily save super-Earths and hot Neptunes around 0.2 AU (the trap however does not seem to survive the end of the collapse phase).
\item Planets,  possibly including those that are quite massive, may be trapped at several locations corresponding to the sublimation lines below 10 AU for any evolution time. In particular, we identify trapping possibilities for gas and ice giants or Neptune-like planets within the few inner astronomical units in the earliest phases.
\item After a few million years, the inner traps may save some close-in super-Earths (we find up to four traps inner to 1.8 AU after 4 Myr).
\item Towards the end of the collapse phase, planets may be trapped further away in the disk at the outer heat transition barrier, regardless of their mass: though they are not required to open a gap there, these planets may still be saved as this heat transition trap appears to be sustainable until the end of the simulation. However massive planets that have not yet opened a gap may be harder to save after a few million years if they have not been trapped earlier, unless they undergo runaway growth as suggested by \citet{crida17}.
\item Gap-opening giant planets can enter type-II migration and eventually escape falling onto the central star by type-I migration, consistently with the expectations of \citet{bitsch15b} that gas giants formed by pebble accretion are born between 10 and 50 AU, and then migrate inward. However, there is no longer a trap leading to a gap-opening zone inside 10 AU.
\end{itemize}

\citetalias{baillie16} already noticed that trapping a planet at the silicate sublimation line is not possible. Though this is also valid here, we may still trap planets close to the star at the inner heat transition barrier.

Towards the end of the collapse phase at around 100,000 years, planets as massive as 200 $M_{\mathrm{\oplus}}$ may be trapped at the sublimation lines, while this maximal mass drops to 60 $M_{\mathrm{\oplus}}$ at the water ice line when the disk has evolved for 1 Myr. While \citet{bitsch15a} showed favorable conditions for trapping planets of up to 50 $M_{\mathrm{\oplus}}$, one has to be careful when trying to compare their work with Figures \ref{mmap200000}-\ref{mmap4000000} here: migration maps should be compared only at similar mass accretion rates. As the model of \citet{bitsch15a}  assumes that the disk is in a steady state with a set mass accretion rate, their figures can only be compared with ours when our disk shows a similar accretion rate. Thus, our 1-million-year-old disk may be compared with their model using a mass accretion rate of $3.5 \, \times \, 10^{-8} \, M_{\odot}/\mathrm{yr}^{-1}$ (their Figure 4). In addition, the maximal masses of the planets that can become trapped have to be compared for a similar outward region (i.e., a similar ice line). Thus, our planets trapped at the water ice line appear to be slightly more massive than theirs (around 37 $M_{\mathrm{\oplus}}$). \citetalias{baillie16} mentioned that part of the difference can be explained by the fact that their model uses a turbulent viscosity $\alpha_{\mathrm{visc}} = 0.0054$, about half of the one used in the present work: this affects the planet mass at which the corotation torque saturates, and therefore the maximal mass of the trapped planets, leading to more massive planets in the present work. In addition, one must remember that the stars and heat diffusion treatments are different between these two models.

Planets of a few Earth masses that are trapped before the end of the collapse phase at the sublimation lines may survive until the end of the disk life while enduring a "trapped migration", outward at first and then inward after the end of the collapse. Some of the less massive planets will survive at the outer heat transition barrier and undergo an outward "trapped migration" before resuming an inward type-I migration. If such a planet gains mass in that latest migration, it may become trapped again at a sublimation line.

Observations of multi-exoplanet systems (Figure 4a of \citet{ogihara15} ) reported periodicity ratio peaks around rational numbers such as 2, $3/2$, $4/3$, $5/4$, and $6/5$, but also wider significant bumps around numbers apparently not related to main resonances, such as 1.8 or 2.2 for instance. After 1 Myr, approximating that they are not planet-mass-dependent and that the modeled disk parameters are representative of the disk population, we report planet traps at 0.55, 0.94, 1.7, and 2.3 AU, meaning that the period ratios of two trapped planets on different orbits would be in the range from 1.57 to 8.55. The  ratio of 1.57 is obtained for the two outermost traps and is consistent with the observed peak slightly above $3/2$. The two innermost traps present a ratio of 2.23, also close to a peak in the observed distribution. Similarly, after 4 Myr, planet traps are located at 0.36, 0.62, 1.12, and 1.65 AU, providing periodicity ratios between 1.79 and 9.81, with 1.79 being due to the two outermost traps and 2.26 owing to the two innermost traps. Again, these two pairs of traps could explain some of the observed nonresonant bumps in the distribution of period ratios of adjacent pairs of planets. In addition, we notice that at both dates, the remaining adjacent pair of traps present a period ratio of 2.43 that could also echo the slightly less important bump in Figure-4a of \citet{ogihara15}.

\citet{steffen15} also observed a peak in the period ratio distribution of Kepler's candidate multiplanet systems around 2.2. Assuming a purely viscously heated steady state inner disk following a power law (equating $T_{m}(r)^{4}$ with the viscous heat flux defined in Section \ref{dem}), we can derive that the midplane temperature follows $T_{m}(r) \propto r^{-5/6}$. From there, we can derive the period ratios of the locations associated with the temperature of the troilite and refractory organics sublimation lines: we evaluate it around 2.33, very close to the observed 2.2 peak. As this 2.2-2.3 ratio is recurrent in our simulations between the two innermost snow-line traps, this could constitute evidence that planets are saved at the snow lines, as suggested by \citet{shvartzvald16} who claim that $55 \%$ of microlensed stars host a snow-line planet.

Though we cannot assume that planets will remain at the trap positions after photo-evaporation and late planet migration, there is a possibility that multiple planets trapped at the sublimation lines around the same star could present period ratios close to the observed peaks.

\subsection{Trapped migration}
As stated previously and detailed in \citet{lyra10}, a trapped planet is likely to remain trapped unless its growth makes it more massive than the maximum mass of the trap. However, these traps do not remain at the same location: not only do they tend to be slightly further in for more massive planets at a given disk age, but they also drift outward as the disk ages during the collapse phase and inward after the infall is finished. Therefore, a trapped planet can still undergo a slow migration that we call trapped migration. Before 170,000 years, this trapped migration may push planetary embryos quite far out in the disk before they resume an inward migration: a planet at the water ice line could be driven as far as 12 AU before turning back for instance. Similarly, a planet trapped at the heat transition barrier could reach up to 35 AU. It is worth noting that this trapped migration exists independently of the formation of the disk by the collapse of the molecular cloud. However, it can push planets outward for a few hundred thousand years in our present simulations while trapped planets would just drift inward in the MMSN case.

\section{Conclusions and perspectives}
\label{cclpersp}
\subsection{Conclusions}

While previous works relied on an initial disk density profile following the controversial MMSN model, we here address that debate by bringing the initial condition back to the parameters of the molecular cloud at the origin of the star and disk. Therefore, we have a better estimation of the disk age than with the MMSN model, since we now model the disk formation by the collapse of the initial molecular cloud instead of assuming it has already formed and reached a power-law density distribution. In that "collapse" scenario, the disk now presents higher midplane temperatures that move the sublimation lines outward.

Another consequence of that disk-building scenario is that a great diversity of planets can now escape falling onto the star by type-I inward migration: very massive planets may open a gap and enter type-II migration before the end of the collapse phase, while low-mass planets can be trapped even in the very inner parts of the disk at the inner heat transition barrier. Jupiter-like embryos may also be trapped between 4 and 9 AU at 200,000 years, as the heat-transition barriers on one hand, and the sublimation lines associated with dust main components on the other, still act as planet traps (except for the silicate line).

In addition, planetary embryos may undergo a trapped migration: they are dragged by the trap that moves outward during the collapse phase and inward afterwards. Such planets may survive below 1.8 AU after 4 Myr of evolution.

Our modeling of the molecular cloud collapse phase actually confirms that such formed protoplanetary disks can be approximated by evolved "MMSN disks" only a few million years after the collapse phase is over which in our case happens around 170,000 years. In addition, this provides indications on the disk structure and the consequences on the planetary embryos during that formation phase: we can then model a much younger disk than was possible with the power-law structure model inherited from the MMSN concept. In particular, the question of the formation of the first solids at the earliest times of the protoplanetary disk requires that disk formation be taken into account and cannot simply be modeled as the evolution of an MMSN. From these comparisons, we could also establish a correspondence between the timelines of disks, depending on whether they form by cloud collapse or are already formed:  the initial states of these two types of simulations appear to be separated by approximately 2 Myr. In the long term, this will certainly allow us to better adjust the birth lines of disks, stars, and early solids in protoplanetary disks.

\subsection{Perspectives}
\label{persp}

Our model is now able to take into consideration the stellar evolution, the viscous spreading of the disk, and the type-I migration: we have access to the optimal conditions for preventing the fall of planetary embryos and we can model the physical and chemical characteristics of those conditions. We now intend to consider a growth model (similar to \citet{alibert05}, \citet{coleman14} or \citet{bitsch15b}) for these embryos to estimate the influence of the planet mass accretion rate on the likelihood of trapping (and therefore the likelihood of a planet surviving a fatal fall onto its star).

In addition, our model cannot currently take into account the counter reaction of the planets onto the disk, or the interactions between multiple planets. These are limitations of our model that should be addressed in order to model the observed resonant planet configurations.

Though we use a detailed opacity model that accounts for the dust-to-gas ratio, we do not consider the dust flux as possibly independent from the gas flux. For instance, extracting dust radial profiles from \citet{gonzalez17} to update our disk thermal structure might be useful.

Now that we have an evolving star model, it would be very interesting to properly account for photo-evaporation based on evolving stellar characteristics. This would help determine the planet distribution at the end of the disk phase.

Finally, we expressed the limits of our stellar model: the use of pre-calculated tables of constant mass star evolution could be improved to account for the accretion luminosity, especially during the collapse phase. One way to properly do that could be to couple a disk-evolution code with a stellar-growth code that would use the disk accretion rate onto the star as an input. Observations of accretion luminosity could then provide initial parameters for modeling the radial profile of potential disks that have not yet been characterized.

\begin{acknowledgements}
The author also thanks Dominic Macdonald for valuable suggestions that improved the quality of the manuscript significantly. This work was supported by the Conseil Scientifique de l'Observatoire de Paris and the Centre National d'Études Spatiales. We also thank the referees for detailed and constructive comments which improved the quality of the paper.
\end{acknowledgements}

\appendix
\section{Lindblad torques}
\label{torquelin}
Assuming the disk to be adiabatic and at thermal equilibrium, the total Lindblad torque exerted by a 2D laminar disk in the absence of self-gravity on a circular planet can be estimated using the formulas detailed in \citet{paarp08}. As thermal diffusion slightly affects the wave propagation velocity, \citet{paar11} defined an effective index $\gamma_{\mathrm{eff}}$ to account for the fact that this velocity is comprised between the isothermal sound speed (maximum thermal diffusion) and the adiabatic sound speed (no thermal diffusion). Therefore, this effective index replaces the $\gamma$-index previously used in the formulas of \citet{paarp08} :

\begin{equation}
\gamma_{\mathrm{eff}} = \frac{2 Q \gamma}{\gamma Q + \frac{1}{2} \sqrt{2 \sqrt{(\gamma^{2}Q^{2}+1)^{2} + 16Q^{2}(\gamma - 1)} + 2 \gamma^{2}Q^{2} - 2}}
,\end{equation}
where $Q$ accounts for the thermal diffusion:
\begin{equation}
Q = \frac{2 \chi_{\mathrm{P}} r_{\mathrm{P}}}{3 h_{pr}^{3}(r_{\mathrm{P}}) \Omega_{\mathrm{P}}},
\end{equation}
and $\chi_{\mathrm{P}}$ is the thermal conductivity at the planet location
\begin{equation}
\chi_{\mathrm{P}} = \frac{16 \gamma (\gamma - 1) \sigma T_{\mathrm{P}}^{4}}{3 \kappa_{\mathrm{P}} \rho_{\mathrm{P}}^{2} h_{pr}^{2}(r_{\mathrm{P}}) \Omega_{\mathrm{P}}}
,\end{equation}
with $\rho_{\mathrm{P}}$ the density, $\kappa_{\mathrm{P}}$ the Rosseland mean opacity, and $\Omega_{\mathrm{P}} = \Omega(r_{\mathrm{P}})$ the Keplerian angular velocity at the planet position in the disk. The 16 factor is a correction introduced by \citet{bitsch11}.

In the isothermal case, $\gamma_{\mathrm{eff}} = 1$, whereas in the adiabatic case, $\gamma_{\mathrm{eff}} = \gamma = 1.4$. The Lindblad torque subsequently becomes

\begin{equation}
\label{gammalin}
\gamma_{\mathrm{eff}} \frac{\Gamma_{\mathrm{Lindblad}}}{\Gamma_{0}(r_{\mathrm{P}})} = - \left(2.5 \, - 1.7 \frac{\partial \ln T}{\partial \ln r} + 0.1 \frac{\partial \ln \Sigma}{\partial \ln r} \right)_{r_{\mathrm{P}}},
\end{equation}
with $q = \frac{M_{planet}}{M_{*}}$ which is the mass ratio of the planet to the star,

\begin{equation}
\label{gamma0}
\Gamma_{0}(r_{\mathrm{P}}) = \left(\frac{q}{h}\right)^{2} \, \Sigma(r_{\mathrm{P}}) \, r_{\mathrm{P}}^{4} \, \Omega_{\mathrm{P}}^{2},
\end{equation}
and $h=\frac{h_{\mathrm{pr}}(r_{\mathrm{P}})}{r_{\mathrm{P}}}$.

\section{Corotation torques}
\label{torquecor}
We estimate the corotation torque by considering the barotropic and entropic contributions separately, both of which contain linear and nonlinear parts. The barotropic was initially studied in \citet{ward91} and \citet{masset01} and detailed in \citet{tanaka02} as a torque arising from the density gradient, while the entropic corotation torque was expressed by \citet{baruteau08} in the case of an adiabatic disk: the horseshoe region presents hotter material in the inner part than in the outer part, inducing, after a U-turn, an excess of mass leading the planet and a deficit of mass trailing behind the planet. This drives angular momentum exchange between the disk and the planet.

For low-enough viscosities ($\alpha_{\mathrm{visc}} < 0.1$), \citet{paarp09b} showed that the corotation torques are mostly non-linear, due to the horseshoe drag caused by the interaction of the planet with the gas in its vicinity \citep{ward91}. As the horseshoe region is closed, it contains a limited amount of angular momentum and therefore is prone to saturation which cancels the horseshoe contributions to the corotation torque.

\citet{paar10} described the fully unsaturated horseshoe drag expressions for both entropic and barotropic (or vortensity) terms. Using the gravitational softening $b=0.4 \, h_{\mathrm{pr}}$ also used in \citet{bitsch11} and \citet{bitsch14}, the horseshoe drag torques read:

\begin{eqnarray}
\label{gammahsentro} \gamma_{\mathrm{eff}} \frac{\Gamma_{\mathrm{hs,entro}}}{\Gamma_{0}(r_{\mathrm{P}})} &=& \frac{7.9}{\gamma_{\mathrm{eff}}} \, \left(-\frac{\partial \ln T}{\partial \ln r} + (\gamma_{\mathrm{eff}}-1) \frac{\partial \ln \Sigma}{\partial \ln r} \right)_{r_{\mathrm{P}}},\\
\label{gammahsbaro} \gamma_{\mathrm{eff}} \frac{\Gamma_{\mathrm{hs,baro}}}{\Gamma_{0}(r_{\mathrm{P}})} &=& 1.1 \left(\frac{\partial \ln \Sigma}{\partial \ln r} + \frac{3}{2}\right)_{r_{\mathrm{P}}}
.\end{eqnarray}

In the general case (including possibly saturation), the total corotation torque is the sum of the barotropic and entropic contributions:
\begin{equation}
\label{gammacor2}
\Gamma_{\mathrm{corotation}} = \Gamma_{\mathrm{c,baro}} + \Gamma_{\mathrm{c,entro}}
,\end{equation}
each of these contributions including a combination of nonlinear (Equations \ref{gammahsentro}-\ref{gammahsbaro}) and linear parts (Equations \ref{gammalinentro}-\ref{gammalinbaro}). The fully unsaturated linear expressions are

\begin{eqnarray}
\label{gammalinentro} \gamma_{\mathrm{eff}} \frac{\Gamma_{\mathrm{lin,entro}}}{\Gamma_{0}(r_{\mathrm{P}})} &=& \left(2.2 - \frac{1.4}{\gamma_{\mathrm{eff}}}\right) \, \left(-\frac{\partial \ln T}{\partial \ln r} + (\gamma_{\mathrm{eff}}-1) \frac{\partial \ln \Sigma}{\partial \ln r} \right)_{r_{\mathrm{P}}},\\
\label{gammalinbaro} \gamma_{\mathrm{eff}} \frac{\Gamma_{\mathrm{lin,baro}}}{\Gamma_{0}(r_{\mathrm{P}})} &=& 0.7 \left(\frac{\partial \ln \Sigma}{\partial \ln r} + \frac{3}{2}\right)_{r_{\mathrm{P}}}
.\end{eqnarray}

Thus, \citet{paar11} expressed the barotropic and entropic torques accounting for the saturation effects as follows.
\begin{eqnarray}
\label{gammacentro}
\Gamma_{\mathrm{c,entro}} &=& F(p_{\nu}) F(p_{\chi}) \sqrt{G(p_{\nu}) G(p_{\chi})} \, \Gamma_{\mathrm{hs,entro}}  \nonumber\\
&& +\,  \sqrt{(1 - K(p_{\nu}))(1 - K(p_{\chi}))} \, \Gamma_{\mathrm{lin,entro}},\\
\label{gammacbaro} \Gamma_{\mathrm{c,baro}} &=& F(p_{\nu}) G(p_{\nu}) \, \Gamma_{\mathrm{hs,baro}} \, + \, (1 - K(p_{\nu})) \, \Gamma_{\mathrm{lin,baro}}
,\end{eqnarray}

where the function $F(p)$ governs the saturation
\begin{equation}
F(p) = \frac{1}{1+(p/1.3)^{2}},
\end{equation}

and the functions $G(p)$ and $K(p)$ govern the cut-off at high viscosity:
\begin{equation}
\label{GG}
G(p) = \left\lbrace
\begin{array}{ccc}
\frac{16}{25} \left( \frac{45 \pi}{8}\right)^{3/4} p^{3/2}  & \mbox{for} & p < \sqrt{\frac{8}{45 \pi}}\\
1 - \frac{9}{25} \left( \frac{8}{45 \pi}\right)^{4/3} p^{-8/3} & \mbox{for} & p \geq \sqrt{\frac{8}{45 \pi}}
\end{array}\right.
,\end{equation}

\begin{equation}
\label{KK}
K(p) = \left\lbrace
\begin{array}{ccc}
\frac{16}{25} \left( \frac{45 \pi}{28}\right)^{3/4} p^{3/2}  & \mbox{for} & p < \sqrt{\frac{28}{45 \pi}}\\
1 - \frac{9}{25} \left( \frac{28}{45 \pi}\right)^{4/3} p^{-8/3} & \mbox{for} & p \geq \sqrt{\frac{28}{45 \pi}}
\end{array}\right.
,\end{equation}
with $p_{\nu}$ being the saturation parameter related to viscosity and $p_{\chi}$ the saturation parameter associated with thermal diffusion:
\begin{eqnarray}
p_{\nu} &=& \frac{2}{3} \, \sqrt{\frac{r_{\mathrm{P}}^{2}\Omega_{\mathrm{P}}x_{s}^{3}}{2 \pi \nu_{\mathrm{P}}}},\\
p_{\chi} &=& \sqrt{\frac{r_{\mathrm{P}}^{2}\Omega_{\mathrm{P}}x_{s}^{3}}{2 \pi \chi_{\mathrm{P}}}},
\end{eqnarray}
where $\nu_{\mathrm{P}}$ and $\chi_{\mathrm{P}}$ are the kinematic viscosity and thermal conductivity at the planet position, and $x_{s}$ is the half width of the horseshoe.
\begin{equation}
\label{xs}
x_{s} = \frac{1.1}{\gamma_{\mathrm{eff}}^{1/4}} \, {\left( \frac{0.4}{\epsilon / h}\right)}^{1/4} \sqrt{\frac{q}{h}}
,\end{equation}
where the smoothing length is $\epsilon / h = b / h_{pr} = 0.4$.

The various contributions of the corotation torque are strongly sensitive to the temperature and surface mass density gradients. Both the corotation and Lindblad torques scale also with $M_{\mathrm{P}}^2$ through $\Gamma_{0}$. The corotation torque also varies with the mass of the planet through the half-width of the horseshoe region; the latter determining the saturation and cut-off coefficients.

\bibliography{bibliography}

\begin{thebibliography}{79}
\expandafter\ifx\csname natexlab\endcsname\relax\def\natexlab#1{#1}\fi

\bibitem[{{Alexander} \& {Armitage}(2007)}]{alexander07}
{Alexander}, R.~D. \& {Armitage}, P.~J. 2007, \mnras, 375, 500

\bibitem[{{Alexander} \& {Armitage}(2009)}]{alexander09}
{Alexander}, R.~D. \& {Armitage}, P.~J. 2009, \apj, 704, 989

\bibitem[{{Alibert} {et~al.}(2005){Alibert}, {Mordasini}, {Benz}, \&
  {Winisdoerffer}}]{alibert05}
{Alibert}, Y., {Mordasini}, C., {Benz}, W., \& {Winisdoerffer}, C. 2005, \aap,
  434, 343

\bibitem[{{Artymowicz}(1993)}]{arty93}
{Artymowicz}, P. 1993, \apj, 419, 155

\bibitem[{{Bailli{\'e}} \& {Charnoz}(2014{\natexlab{a}})}]{baillie14iaus}
{Bailli{\'e}}, K. \& {Charnoz}, S. 2014{\natexlab{a}}, in IAU Symposium, Vol.
  299, Exploring the Formation and Evolution of Planetary Systems, ed.
  M.~{Booth}, B.~C. {Matthews}, \& J.~R. {Graham}, 374--375

\bibitem[{{Bailli{\'e}} \& {Charnoz}(2014{\natexlab{b}})}]{baillie14}
{Bailli{\'e}}, K. \& {Charnoz}, S. 2014{\natexlab{b}}, \apj, 786, 35

\bibitem[{{Bailli{\'e}} {et~al.}(2015){Bailli{\'e}}, {Charnoz}, \&
  {Pantin}}]{baillie15}
{Bailli{\'e}}, K., {Charnoz}, S., \& {Pantin}, E. 2015, \aap, 577, A65

\bibitem[{{Bailli{\'e}} {et~al.}(2016){Bailli{\'e}}, {Charnoz}, \&
  {Pantin}}]{baillie16}
{Bailli{\'e}}, K., {Charnoz}, S., \& {Pantin}, E. 2016, \aap, 590, A60

\bibitem[{{Barranco} \& {Goodman}(1998)}]{barranco98}
{Barranco}, J.~A. \& {Goodman}, A.~A. 1998, \apj, 504, 207

\bibitem[{{Baruteau} \& {Masset}(2008)}]{baruteau08}
{Baruteau}, C. \& {Masset}, F. 2008, \apj, 672, 1054

\bibitem[{{Beckwith} \& {Sargent}(1996)}]{beckwith96}
{Beckwith}, S.~V.~W. \& {Sargent}, A.~I. 1996, \nat, 383, 139

\bibitem[{{Bell} \& {Lin}(1994)}]{bell94}
{Bell}, K.~R. \& {Lin}, D.~N.~C. 1994, \apj, 427, 987

\bibitem[{{Bitsch} {et~al.}(2013){Bitsch}, {Crida}, {Morbidelli}, {Kley}, \&
  {Dobbs-Dixon}}]{bitsch13}
{Bitsch}, B., {Crida}, A., {Morbidelli}, A., {Kley}, W., \& {Dobbs-Dixon}, I.
  2013, \aap, 549, A124

\bibitem[{{Bitsch} {et~al.}(2015{\natexlab{a}}){Bitsch}, {Johansen},
  {Lambrechts}, \& {Morbidelli}}]{bitsch15a}
{Bitsch}, B., {Johansen}, A., {Lambrechts}, M., \& {Morbidelli}, A.
  2015{\natexlab{a}}, \aap, 575, A28

\bibitem[{{Bitsch} \& {Kley}(2011)}]{bitsch11}
{Bitsch}, B. \& {Kley}, W. 2011, \aap, 536, A77

\bibitem[{{Bitsch} {et~al.}(2015{\natexlab{b}}){Bitsch}, {Lambrechts}, \&
  {Johansen}}]{bitsch15b}
{Bitsch}, B., {Lambrechts}, M., \& {Johansen}, A. 2015{\natexlab{b}}, \aap,
  582, A112

\bibitem[{{Bitsch} {et~al.}(2014){Bitsch}, {Morbidelli}, {Lega}, \&
  {Crida}}]{bitsch14}
{Bitsch}, B., {Morbidelli}, A., {Lega}, E., \& {Crida}, A. 2014, \aap, 564,
  A135

\bibitem[{{Calvet} {et~al.}(1991){Calvet}, {Patino}, {Magris}, \&
  {D'Alessio}}]{calvet91}
{Calvet}, N., {Patino}, A., {Magris}, G.~C., \& {D'Alessio}, P. 1991, \apj,
  380, 617

\bibitem[{{Charnoz} {et~al.}(2015){Charnoz}, {Al{\'e}on}, {Chaumard},
  {Bailli{\'e}}, \& {Taillifet}}]{charnoz15}
{Charnoz}, S., {Al{\'e}on}, J., {Chaumard}, N., {Bailli{\'e}}, K., \&
  {Taillifet}, E. 2015, \icarus, 252, 440

\bibitem[{{Chiang} \& {Goldreich}(1997)}]{chiang97}
{Chiang}, E.~I. \& {Goldreich}, P. 1997, \apj, 490, 368

\bibitem[{{Coleman} \& {Nelson}(2014)}]{coleman14}
{Coleman}, G.~A.~L. \& {Nelson}, R.~P. 2014, \mnras, 445, 479

\bibitem[{{Crida}(2009)}]{crida09}
{Crida}, A. 2009, \apj, 698, 606

\bibitem[{{Crida} \& {Bitsch}(2017)}]{crida17}
{Crida}, A. \& {Bitsch}, B. 2017, \icarus, 285, 145

\bibitem[{{Crida} {et~al.}(2006){Crida}, {Morbidelli}, \& {Masset}}]{crida06}
{Crida}, A., {Morbidelli}, A., \& {Masset}, F. 2006, \icarus, 181, 587

\bibitem[{{Dapp} \& {Basu}(2010)}]{dapp10}
{Dapp}, W.~B. \& {Basu}, S. 2010, \aap, 521, L56

\bibitem[{{Flock} {et~al.}(2016){Flock}, {Fromang}, {Turner}, \&
  {Benisty}}]{flock16}
{Flock}, M., {Fromang}, S., {Turner}, N.~J., \& {Benisty}, M. 2016, \apj, 827,
  144

\bibitem[{{Flock} {et~al.}(2017){Flock}, {Fromang}, {Turner}, \&
  {Benisty}}]{flock17}
{Flock}, M., {Fromang}, S., {Turner}, N.~J., \& {Benisty}, M. 2017, \apj, 835,
  230

\bibitem[{{Font} {et~al.}(2004){Font}, {McCarthy}, {Johnstone}, \&
  {Ballantyne}}]{font04}
{Font}, A.~S., {McCarthy}, I.~G., {Johnstone}, D., \& {Ballantyne}, D.~R. 2004,
  \apj, 607, 890

\bibitem[{{Goldreich} \& {Tremaine}(1979)}]{goldreich79}
{Goldreich}, P. \& {Tremaine}, S. 1979, \apj, 233, 857

\bibitem[{{Gonzalez} {et~al.}(2017){Gonzalez}, {Laibe}, \&
  {Maddison}}]{gonzalez17}
{Gonzalez}, J.-F., {Laibe}, G., \& {Maddison}, S.~T. 2017, \mnras

\bibitem[{{Goodman} {et~al.}(1993){Goodman}, {Benson}, {Fuller}, \&
  {Myers}}]{goodman93}
{Goodman}, A.~A., {Benson}, P.~J., {Fuller}, G.~A., \& {Myers}, P.~C. 1993,
  \apj, 406, 528

\bibitem[{{Hartmann} {et~al.}(1998){Hartmann}, {Calvet}, {Gullbring}, \&
  {D'Alessio}}]{hartmann98}
{Hartmann}, L., {Calvet}, N., {Gullbring}, E., \& {D'Alessio}, P. 1998, \apj,
  495, 385

\bibitem[{{Hasegawa} \& {Pudritz}(2011)}]{hasegawa111}
{Hasegawa}, Y. \& {Pudritz}, R.~E. 2011, \mnras, 413, 286

\bibitem[{{Hayashi}(1981)}]{hayashi81}
{Hayashi}, C. 1981, Progress of Theoretical Physics Supplement, 70, 35

\bibitem[{{Helling} {et~al.}(2000){Helling}, {Winters}, \&
  {Sedlmayr}}]{helling00}
{Helling}, C., {Winters}, J.~M., \& {Sedlmayr}, E. 2000, \aap, 358, 651

\bibitem[{{Hennebelle} \& {Fromang}(2008)}]{hennebellef08}
{Hennebelle}, P. \& {Fromang}, S. 2008, \aap, 477, 9

\bibitem[{{Hueso} \& {Guillot}(2005)}]{hueso05}
{Hueso}, R. \& {Guillot}, T. 2005, \aap, 442, 703

\bibitem[{{Jang-Condell} \& {Sasselov}(2005)}]{jc05}
{Jang-Condell}, H. \& {Sasselov}, D.~D. 2005, \apj, 619, 1123

\bibitem[{{Joos} {et~al.}(2012){Joos}, {Hennebelle}, \& {Ciardi}}]{joos12}
{Joos}, M., {Hennebelle}, P., \& {Ciardi}, A. 2012, \aap, 543, A128

\bibitem[{{Li} {et~al.}(2014){Li}, {Banerjee}, {Pudritz}, {J{\o}rgensen},
  {Shang}, {Krasnopolsky}, \& {Maury}}]{li14}
{Li}, Z.-Y., {Banerjee}, R., {Pudritz}, R.~E., {et~al.} 2014, Protostars and
  Planets VI, 173

\bibitem[{{Lodato}(2008)}]{lodato08}
{Lodato}, G. 2008, \nar, 52, 21

\bibitem[{{Lynden-Bell} \& {Pringle}(1974)}]{lyndenbellpringle74}
{Lynden-Bell}, D. \& {Pringle}, J.~E. 1974, \mnras, 168, 603

\bibitem[{{Lyra} {et~al.}(2010){Lyra}, {Paardekooper}, \& {Mac Low}}]{lyra10}
{Lyra}, W., {Paardekooper}, S.-J., \& {Mac Low}, M.-M. 2010, \apjl, 715, L68

\bibitem[{{Marques} {et~al.}(2013){Marques}, {Goupil}, {Lebreton}, {Talon},
  {Palacios}, {Belkacem}, {Ouazzani}, {Mosser}, {Moya}, {Morel}, {Pichon},
  {Mathis}, {Zahn}, {Turck-Chi{\`e}ze}, \& {Nghiem}}]{marques13}
{Marques}, J.~P., {Goupil}, M.~J., {Lebreton}, Y., {et~al.} 2013, \aap, 549,
  A74

\bibitem[{{Masset}(2001)}]{masset01}
{Masset}, F.~S. 2001, \apj, 558, 453

\bibitem[{{Masset} {et~al.}(2006){Masset}, {Morbidelli}, {Crida}, \&
  {Ferreira}}]{masset06b}
{Masset}, F.~S., {Morbidelli}, A., {Crida}, A., \& {Ferreira}, J. 2006, \apj,
  642, 478

\bibitem[{{Masunaga} \& {Inutsuka}(1999)}]{masunaga99}
{Masunaga}, H. \& {Inutsuka}, S.-i. 1999, \apj, 510, 822

\bibitem[{{Masunaga} {et~al.}(1998){Masunaga}, {Miyama}, \&
  {Inutsuka}}]{masunaga98}
{Masunaga}, H., {Miyama}, S.~M., \& {Inutsuka}, S.-i. 1998, \apj, 495, 346

\bibitem[{{Maury} {et~al.}(2010){Maury}, {Andr{\'e}}, {Hennebelle}, {Motte},
  {Stamatellos}, {Bate}, {Belloche}, {Duch{\^e}ne}, \& {Whitworth}}]{maury10}
{Maury}, A.~J., {Andr{\'e}}, P., {Hennebelle}, P., {et~al.} 2010, \aap, 512,
  A40

\bibitem[{{Mellon} \& {Li}(2008)}]{mellon08}
{Mellon}, R.~R. \& {Li}, Z.-Y. 2008, \apj, 681, 1356

\bibitem[{{Menou} \& {Goodman}(2004)}]{menou04}
{Menou}, K. \& {Goodman}, J. 2004, \apj, 606, 520

\bibitem[{{Morbidelli} {et~al.}(2008){Morbidelli}, {Crida}, {Masset}, \&
  {Nelson}}]{morbidelli08}
{Morbidelli}, A., {Crida}, A., {Masset}, F., \& {Nelson}, R.~P. 2008, \aap,
  478, 929

\bibitem[{{Morel}(1997)}]{morel97}
{Morel}, P. 1997, \aaps, 124, 597

\bibitem[{{Morel} \& {Lebreton}(2008)}]{morel08}
{Morel}, P. \& {Lebreton}, Y. 2008, \apss, 316, 61

\bibitem[{{Ogihara} {et~al.}(2015){Ogihara}, {Morbidelli}, \&
  {Guillot}}]{ogihara15}
{Ogihara}, M., {Morbidelli}, A., \& {Guillot}, T. 2015, \aap, 578, A36

\bibitem[{{Owen} {et~al.}(2010){Owen}, {Ercolano}, {Clarke}, \&
  {Alexander}}]{owen10}
{Owen}, J.~E., {Ercolano}, B., {Clarke}, C.~J., \& {Alexander}, R.~D. 2010,
  \mnras, 401, 1415

\bibitem[{{Paardekooper} {et~al.}(2010){Paardekooper}, {Baruteau}, {Crida}, \&
  {Kley}}]{paar10}
{Paardekooper}, S.-J., {Baruteau}, C., {Crida}, A., \& {Kley}, W. 2010, \mnras,
  401, 1950

\bibitem[{{Paardekooper} {et~al.}(2011){Paardekooper}, {Baruteau}, \&
  {Kley}}]{paar11}
{Paardekooper}, S.-J., {Baruteau}, C., \& {Kley}, W. 2011, \mnras, 410, 293

\bibitem[{{Paardekooper} \& {Papaloizou}(2008)}]{paarp08}
{Paardekooper}, S.-J. \& {Papaloizou}, J.~C.~B. 2008, \aap, 485, 877

\bibitem[{{Paardekooper} \& {Papaloizou}(2009{\natexlab{a}})}]{paarp09a}
{Paardekooper}, S.-J. \& {Papaloizou}, J.~C.~B. 2009{\natexlab{a}}, \mnras,
  394, 2283

\bibitem[{{Paardekooper} \& {Papaloizou}(2009{\natexlab{b}})}]{paarp09b}
{Paardekooper}, S.-J. \& {Papaloizou}, J.~C.~B. 2009{\natexlab{b}}, \mnras,
  394, 2297

\bibitem[{{Palla} \& {Stahler}(1990)}]{palla90}
{Palla}, F. \& {Stahler}, S.~W. 1990, \apjl, 360, L47

\bibitem[{{Piau} {et~al.}(2011){Piau}, {Kervella}, {Dib}, \&
  {Hauschildt}}]{piau11}
{Piau}, L., {Kervella}, P., {Dib}, S., \& {Hauschildt}, P. 2011, \aap, 526,
  A100

\bibitem[{{Pollack} {et~al.}(1996){Pollack}, {Hubickyj}, {Bodenheimer},
  {Lissauer}, {Podolak}, \& {Greenzweig}}]{pollack96}
{Pollack}, J.~B., {Hubickyj}, O., {Bodenheimer}, P., {et~al.} 1996, \icarus,
  124, 62

\bibitem[{{Pringle}(1981)}]{pringle81}
{Pringle}, J.~E. 1981, \araa, 19, 137

\bibitem[{{Seifried} {et~al.}(2013){Seifried}, {Banerjee}, {Pudritz}, \&
  {Klessen}}]{seifried13}
{Seifried}, D., {Banerjee}, R., {Pudritz}, R.~E., \& {Klessen}, R.~S. 2013,
  \mnras, 432, 3320

\bibitem[{{Semenov} {et~al.}(2003){Semenov}, {Henning}, {Helling}, {Ilgner}, \&
  {Sedlmayr}}]{semenov03}
{Semenov}, D., {Henning}, T., {Helling}, C., {Ilgner}, M., \& {Sedlmayr}, E.
  2003, \aap, 410, 611

\bibitem[{{Shakura} \& {Sunyaev}(1973)}]{shakura73}
{Shakura}, N.~I. \& {Sunyaev}, R.~A. 1973, \aap, 24, 337

\bibitem[{{Shu}(1977)}]{shu77}
{Shu}, F.~H. 1977, \apj, 214, 488

\bibitem[{{Shvartzvald} {et~al.}(2016){Shvartzvald}, {Maoz}, {Udalski}, {Sumi},
  {Friedmann}, {Kaspi}, {Poleski}, {Szyma{\'n}ski}, {Skowron}, {Koz{\l}owski},
  {Wyrzykowski}, {Mr{\'o}z}, {Pietrukowicz}, {Pietrzy{\'n}ski},
  {Soszy{\'n}ski}, {Ulaczyk}, {Abe}, {Barry}, {Bennett}, {Bhattacharya},
  {Bond}, {Freeman}, {Inayama}, {Itow}, {Koshimoto}, {Ling}, {Masuda}, {Fukui},
  {Matsubara}, {Muraki}, {Ohnishi}, {Rattenbury}, {Saito}, {Sullivan},
  {Suzuki}, {Tristram}, {Wakiyama}, \& {Yonehara}}]{shvartzvald16}
{Shvartzvald}, Y., {Maoz}, D., {Udalski}, A., {et~al.} 2016, \mnras, 457, 4089

\bibitem[{{Steffen} \& {Hwang}(2015)}]{steffen15}
{Steffen}, J.~H. \& {Hwang}, J.~A. 2015, \mnras, 448, 1956

\bibitem[{{Tanaka} {et~al.}(2002){Tanaka}, {Takeuchi}, \& {Ward}}]{tanaka02}
{Tanaka}, H., {Takeuchi}, T., \& {Ward}, W.~R. 2002, \apj, 565, 1257

\bibitem[{{Tobin} {et~al.}(2015){Tobin}, {Looney}, {Wilner}, {Kwon},
  {Chandler}, {Bourke}, {Loinard}, {Chiang}, {Schnee}, \& {Chen}}]{tobin15}
{Tobin}, J.~J., {Looney}, L.~W., {Wilner}, D.~J., {et~al.} 2015, \apj, 805, 125

\bibitem[{{van Dishoeck} {et~al.}(1993){van Dishoeck}, {Blake}, {Draine}, \&
  {Lunine}}]{vandishoeck1993}
{van Dishoeck}, E.~F., {Blake}, G.~A., {Draine}, B.~T., \& {Lunine}, J.~I.
  1993, in Protostars and Planets III, ed. E.~H. {Levy} \& J.~I. {Lunine},
  163--241

\bibitem[{{Ward}(1988)}]{ward88}
{Ward}, W.~R. 1988, \icarus, 73, 330

\bibitem[{{Ward}(1991)}]{ward91}
{Ward}, W.~R. 1991, in Lunar and Planetary Science Conference, Vol.~22, Lunar
  and Planetary Science Conference, 1463

\bibitem[{{Ward}(1997)}]{ward97}
{Ward}, W.~R. 1997, \icarus, 126, 261

\bibitem[{{Weidenschilling}(1977)}]{weiden77}
{Weidenschilling}, S.~J. 1977, \apss, 51, 153

\bibitem[{{Yang} \& {Ciesla}(2012)}]{yang12}
{Yang}, L. \& {Ciesla}, F.~J. 2012, Meteoritics and Planetary Science, 47, 99

\end{thebibliography}
\end{document}